\documentclass[english]{article}
\usepackage{times}
\usepackage[T1]{fontenc}
\usepackage[latin1]{inputenc}
\usepackage{geometry}
\usepackage{rotating}
\usepackage{color}
\usepackage{graphicx}
\usepackage{setspace}
\usepackage{amssymb}
\usepackage[authoryear]{natbib}

\makeatletter

\providecommand{\tabularnewline}{\\}

\usepackage{babel}
\makeatother
\begin{document}

\large

\title{The Fixation Probability of Rare Mutators in Finite Asexual Populations\\
}

\author{~\\
\\
\\
\\
 \\
C. Scott Wylie$^{1}$, Cheol-Min Ghim$^{2}$, David Kessler$^{3}$,
Herbert Levine$^{1}$}

\date{~ }

\maketitle
~\\
\\
\\
\\
\\
\\
\\
\\
\\
\\
$^{1}$Center For Theoretical Biological Physics, University of California, San Diego, USA

\noindent $^{2}$Lawrence Livermore National Laboratories, Livermore, CA, USA

\noindent $^{3}$Department of Physics, Bar-Ilan University, Ramat-Gan, Israel

\clearpage

Running Head: Mutator Fixation Probability\\

Keywords: mutator, genetic instability, random drift, indirect selection, mutation rate evolution \\

Corresponding author:

C. Scott Wylie

Center for Theoretical Biological Physics 

University of California, San Diego 

La Jolla, CA 92093

USA

Phone 858-534-5731

Fax 858-534-7697

Email: cwylie@physics.ucsd.edu

\clearpage

\begin{abstract}
{\large A mutator is an allele that increases the mutation rate throughout the genome by disrupting some aspect of DNA replication or repair.  Mutators that increase the mutation rate by the order of 100 fold have been observed to spontaneously emerge and achieve high frequencies in natural populations and in long-term laboratory evolution experiments with \textit{E. coli}.   In principle, the fixation of mutator alleles is limited by (i) competition with mutations in wild-type backgrounds, (ii) additional deleterious mutational load, and (iii) random genetic drift.  Using a multiple locus model and employing both simulation and analytic methods, we investigate the effects of these three factors on the fixation probability $P_{fix}$ of an initially rare mutator as a function of population size $N$, beneficial and deleterious mutation rates, and the strength of mutations $s$.  Our diffusion based approximation for $P_{fix}$ successfully captures effects (ii) and (iii) when selection is fast compared to mutation ($\mu/s \ll 1$).  This enables us to predict the conditions under which mutators will be evolutionarily favored.  Surprisingly, our simulations show that effect (i) is typically small for strong-effect mutators.  Our results agree semi-quantitatively with existing laboratory evolution experiments and suggest future experimental directions. }{\large \par}
\end{abstract}
\clearpage

The most evolutionarily important characteristic that an individual inherits from its parents is the average number of offspring that it will leave in the next generation, i.e. its fitness.  But, is fitness the \textit{only} evolutionarily relevant heritable trait?  The ultimate fate of an individual depends not only on its immediate properties, but  on those of its entire lineage of descendants.   Therefore, the genetic system that shapes the statistical properties of this lineage is also an evolutionarily relevant, selectable trait.  

In this article we study one such property, namely a globally elevated mutation rate.  In practice this property is inherited via a mutated copy of a gene, called a mutator allele, involved in DNA copy or repair.  We ask the following basic question: \textit{What is the fixation probability of an initially rare mutator?}  This is a generalization of the classic population genetic calculation for the fixation probability of a static mutant with selection coefficient $s$ \citep{fisher1930gtn}.  If the fixation probability of a mutator allele differs from that of a neutral one (i.e. $1/N$), then the average mutation rate of the population will be under selective pressure.

The selective forces acting on mutators is not purely a theoretical issue. Natural populations quite often contain a mixture of wild-type and mutator strains \citep{leclerc1996hmf,leclerc2000pss,giraud2001cab,matic1997hvm,delcampo2004psa,bjorkholm2004mfa,oliver2000hfh,prunier2003hrm,richardson2002mcn,watson2004hhi}.  Furthermore, the somatic tissues of multicellular sexual organisms comprise populations of asexually reproducing cells possessing opportunities for an increased growth rate.  Correspondingly, tumoregenesis has been associated with mutator alleles \citep{loeb1991mpm}.  Even more strikingly, laboratory-scale evolution experiments \citep {sniegowski1997ehm,mao1997pmc,treffers1954fom,miyake1960mfs} have resulted in examples of spontaneous mutator fixation.  Several experimental studies \citep{giraud2001cab,chao1983cbh,labat2005mpc,shaver2002fea,mao1997pmc} indicate that mutators achieve fixation because of the adaptive mutations they generate and not because of any intrinsic fitness advantage.  Thus, selection on mutator alleles occurs via an indirect mechanism.   One of the goals of our work is to make  semi-quantitative contact between our model of indirect selection and the existing data of mutator fixation in laboratory experiments.  

The evolution of mutation rate is a problem that dates back to the 1930's. The general issue was articulated by \cite{sturtevant1937eee}, and important theoretical contributions date back to \cite{kimura1967eas} and \cite{leighorig}. Theoretical studies proliferated during the last decade, and the field is reviewed by \cite{sniegowski2000emr} and also by \cite{maticrev}.  Given the abundance of existing theoretical articles, it is critical to understand how our work relates to and improves upon this body of literature.  We address this issue in detail in the Discussion section.  For now, we merely provide a brief sketch.
First, we neglect the complicating influences of recombination and environmental fluctuations.  This allows for a direct and comparatively precise treatment of the simplest situation: a strictly asexual population adapting in a constant environment.  Even this simplest scenario has rich and often counterintuitive behavior.  Secondly, our methods naturally treat both strong (e.g. 100 fold) mutators and weak modifiers of mutation rate. Thirdly, unlike most previous work, we combine fully stochastic simulations with an analytic approach.  Our analytic results for weak modifiers are a generalization of previous work by \cite{andre2006emr}, but we find that both approaches often fail to match simulations.  However, our work for strong mutators \textit{does} match simulations over the expected parameter range.  The simulations thus provide vital checks and guidance for the analytic approach.  Conversely, the analytic approach deepens our understanding of mutator fixation and makes predictions in parameter regimes that are computationally inaccessible via simulation.  Finally, unlike previous work, our diffusion based analytic approach captures the effects of random genetic drift.  This not only allows for exploration of regimes where random drift is important, but also a quantitative understanding of when it can be neglected.  

The outline of this article is as follows.  We begin with a heuristic discussion of mutator dynamics.  Next, we construct and simulate a stochastic model of asexual populations that include mutator alleles. We do not explicitly allow for the formation of mutators, merely the competition between mutators and wild-type strains once mutators arise. Afterward, steered by the outcome of simulations, we develop a quantitative understanding of the results of the stochastic simulations.  Although a full mathematical treatment turns out to be intractable, we are able to devise an approximation scheme that captures many features of the simulation results.  We then solve our approximation scheme, both numerically and analytically.  The resulting expressions allow a comparison to the \textit{E. coli} experiments of Lenski and co-workers \citep{sniegowski1997ehm}.    
\\

\begin{center}{HEURISTIC ANALYSIS}\end{center}

\begin{table}
\begin{doublespace}

\caption{\large  \doublespacing
 Commonly used notation. }
\end{doublespace}

\begin{center}\begin{tabular}{ll}
\hline 
Symbol & Usage
\tabularnewline
\hline
\tabularnewline
N            & Total population size
\tabularnewline
$\mu_-$ & Wild-type mutation rate per genome
\tabularnewline
$\mu_+$ & Mutator mutation rate per genome
\tabularnewline
$U$       & Mutation rate into mutator state
\tabularnewline
$L$         & Length of genome
\tabularnewline
$b$        & Number of 1's in genome
\tabularnewline
$\delta$ & Fraction of mutations that are lethal
\tabularnewline
$x$         & Mutator frequency
\tabularnewline
$\bar \mu \equiv (1-x)\mu_- + x \mu_+$  &  Average mutation rate per genome
\tabularnewline
$R \equiv \mu_+/\mu_-  $ & Mutator strength
\tabularnewline
$r \equiv b/L$        &  Growth rate per individual per simulation time-step
\tabularnewline
$s = 1/b$ & Selection coefficient of non-lethal mutation
\tabularnewline
$\alpha \equiv 1-b/L$  & Fraction of 0's in the genome
\tabularnewline
$\alpha_e \equiv \alpha (1-\delta)$ & Fraction of mutations that are beneficial

\tabularnewline
\hline
\end{tabular}\end{center}
\label{param_table}
\end{table}

In this section, we briefly explain the conceptual factors underlying mutator fixation.  The equations in this section should be considered merely as heuristic guides and not formal results.  

\begin{figure}
\includegraphics[width=6in]{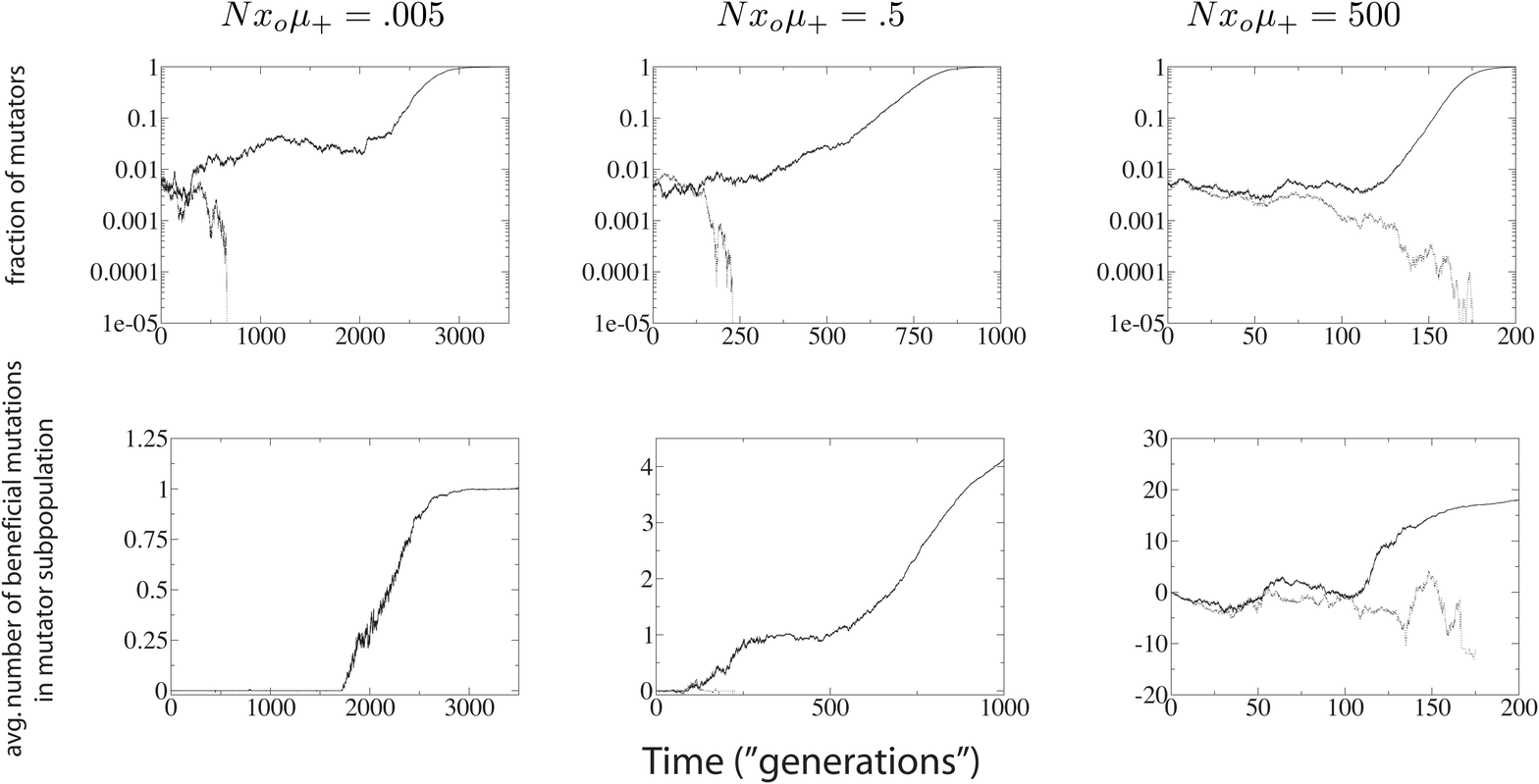}
\caption{\large \doublespacing Some sample runs from simulations where the wild-type mutation rate is zero. The top panels depict the number of mutators in the population vs. $\frac{rt}{N}$, where $r$ is the birth probability per time step which is proportional to the (initial) mean population fitness.  The bottom panels show the average number of beneficial mutations in the mutator subpopulation.  The dark lines resulted in fixation of the mutator allele, whereas the lighter lines resulted in its loss.   When the mutation rate of the mutators ($\mu_+$) is not too large, the mutator hitchhikes to fixation with a single beneficial mutation (left panels).  When $\mu_+$ is larger, many beneficial mutations occur during the fixation process (center and right panels).  Our analytic approximation scheme assumes that the fixation process is \textit{triggered} by merely the first beneficial mutation to survive drift.  Note that in each case the population is always far from the fitness maximum when the mutator achieves fixation since there are 80 possible beneficial mutations.  Parameters are $N=10^5, x_o=.005,\delta = 0,$ wild-type mutation rate $\mu_-=0$, and $\mu_+=10^{-5}$ (left), $\mu_+=10^{-3}$ (center), $\mu_+=1$ (right).  $\alpha=.4,s=1/120$ (initial values).  } 
\label{sample_sims}
\end{figure}

Since mutator alleles do not directly affect fitness, their dynamics must be guided by association with other genes which do have a direct fitness effect.  In asexuals, all loci sharing the same genome with a  sweeping beneficial mutation will also become fixed via ``hitchhiking'' \citep{smith1974hhe}.  Whereas most alleles hitchhike completely passively, the mutator allele plays a somewhat active role in facilitating its own hitchhiking by increasing the probability of a beneficial mutation elsewhere in the genome.  This well known mechanism occurs in our simulations and is evident in Fig.\ref{sample_sims}.

At the same time, the wild-type subpopulation also generates advantageous mutations.  When this occurs, mutators become extinct due to fixation of their counterpart wild-type alleles.  Although the wild-type generates mutations more slowly on a per capita basis, if it vastly outnumbers the mutator subpopulation, then the \textit{total} mutation rate in the wild-type background may be larger.  Along these lines, it is tempting to think of the number of mutators as initially constant, and that the mutator will achieve fixation if and only if it generates a sweeping beneficial mutation before the wild-type background does.  This means that  
\begin{equation}
P_{fix} = x_o\frac{\mu_+}{\bar{\mu}} = \frac{x_o \mu_+}{x_o\mu_+ + (1-x_o)\mu_-}
\label{freq_dependent}
\end{equation} 
where $x_o$ is the initial frequency of mutators and $\mu_+$ ($\mu_-$) is the genome-wide  mutator (wild-type) mutation rate.  This equation has striking qualities.  First, it is independent of the following \textit{prima facie} important parameters: population size $N$, selection coefficient of mutations $s$, and the fraction of mutations which are beneficial versus deleterious.  Secondly, and more subtly, the equation is \textit{explicitly frequency dependent}.  It will turn out that Eq.\ref{freq_dependent} arises as a limiting form of our analytic expression, but does \textit{not} typically match the results of simulations.

In contrast to the frequency dependent Eq.\ref{freq_dependent}, a classic result from population genetics \citep{fisher1930gtn} is the fixation probability of a mutant with a simple selective advantage: 
\begin{equation}
P_{fix} = \frac{1-e^{-Nx_oS}}{1-e^{-NS}}
\label{classical_fix}
\end{equation}
This result holds for haploid populations using Moran process dynamics, and merely requires factors of two in the exponents to handle diploids or Wright-Fisher dynamics.  In Eq.\ref{classical_fix}, $P_{fix}$ depends on the frequency of mutants only via the product $Nx_o$, i.e. the initial \textit{number} of mutants.  Thus, Eqs.\ref{freq_dependent},\ref{classical_fix} scale differently with population size.  The form of Eq.\ref{classical_fix} implies that (when $NS \gg 1$), $P_{fix} \approx 1-e^{-Nx_oS} \approx 1-(1-S)^{Nx_o}$ and we can think of each mutant as an independent ``trial'' with fixation probability $S$.  In other words, if the fraction $x_o$ is kept constant and $N$ is increased, Eq.\ref{freq_dependent} says that $P_{fix}$ should remain unchanged whereas Eq.\ref{classical_fix} says that $P_{fix}$ should increase.  On the other hand, if $Nx_o$ is held constant as $N$ is increased, Eq.\ref{freq_dependent} predicts a decrease in $P_{fix}$ whereas Eq.\ref{classical_fix} predicts that $P_{fix}$ remains unchanged.  Since mutators achieve fixation by hitchhiking with mutations which are themselves governed by Eq.\ref{classical_fix}, perhaps we should \textit{a priori}  view Eq.\ref{freq_dependent} with suspicion.  Indeed, our simulation data and analytic methods will show that mutator fixation is often governed by an equation with the form of Eq.\ref{classical_fix}.

While Eq.\ref{freq_dependent} completely neglects deleterious mutations, they are the basis for another heuristic line of thought.  In any realistic biological population, regardless of how maladapted, deleterious mutations vastly outnumber advantageous ones.   Because of this, upon first thought, one might think that the mutator allele will do more harm than good and therefore be selected against.  Although it is true that an elevated mutation rate will quite likely cause an immediate decrease in the population's mean fitness, evolution does not always act to maximize this quantity.  The situation is understood more clearly in the following game theoretical context.  A beneficial mutation often greatly increases the probability that a lineage will achieve complete evolutionary success by sweeping through the entire population, whereas a deleterious mutation only slightly decreases the low probability of a neutral sweep.  More quantitatively, we can think of the ``payoff'' for a sweeping advantageous mutant as the entire population size $N$.  For this to occur, the mutator must generate a beneficial mutation which must then survive in spite of random drift.  In contrast, the payoff for a deleterious mutant is merely a single individual who is destined to die out with near certainty.  The mutation strategy is favored when its expected payoff is greater than zero, i.e.
\begin{equation}
N\cdot \pi(s) \cdot \mu_{ben} - 1\cdot \mu_{del} > 0
\label{game_theory_trans}
\end{equation}
where $\pi(s)$ is the fixation probability of a simple mutant, given by Eq.\ref{classical_fix} and $\mu_{ben}$ ($\mu_{del}$) are the beneficial and deleterious mutation rates, respectively.  Note that this expression weights beneficial mutations $N\cdot \pi(s)$ times more heavily than deleterious ones, underscoring their asymmetric effects.  Later in this article, we show that Eq.\ref{game_theory_trans} also follows from a rigorous mathematical analysis.

Thus far we have argued that the fate of mutators is in principle limited both by competition with wild-type and by their increased load of deleterious mutants.  Additionally, random genetic drift is commonly a potent force acting on rare subpopulations.  Each mutator begins its existence selectively neutral.  It can be shown that random drift eliminates neutral alleles from the population with a high probability $=1-1/N$, and that the average time taken to do so is merely $\sim \ln(N)$ generations \citep{crow:ipg}.  
Although we cannot write down a ``back of the envelope'' estimate of this effect, we will later derive a formula that fully incorporates random drift and specifies the parameter regimes in which it dominates mutator fixation. 

Our analytic work results in a formula for the mutator fixation probability in terms of simple parameters.  Examining this expression yields a quantitative sense of the relative importance of random drift, deleterious mutations, and beneficial mutations.  This allows us to define ``strong-effect'' and ``weak-effect'' mutator regimes in terms of the model parameters.  In the strong-effect regime, mutations in the wild-type background do not affect mutator success and our analytic approach works well.  In the weak-effect regime, mutations in wild-type backgrounds are predicted to be the dominant influence on mutator fixation.  However, in the case of weak-effect mutators, we will show that our analytic approach, like existing work by \cite{andre2006emr}, typically overestimates the competitive effects of mutations in wild-type backgrounds.  When this is true, Eq.\ref{freq_dependent} provides a poor description of mutator fixation.  We now turn toward a discussion of our stochastic simulations, that provide an invaluable reference to which we compare our analytic work.\\

\begin{center}{DESCRIPTION OF STOCHASTIC SIMULATIONS}\end{center}

We model haploid asexual populations of fixed size $N$ undergoing stochastic processes of birth, death, and mutation. Initially, a fraction $x_o \ll 1$ of the population are mutators and all individuals have the same fitness.  The birth-death-mutation process is iterated until the population consists entirely of either mutators or wild-type.  Transitions between the mutator and wild-type states are not allowed.  We do not model environmental changes explicitly, thereby assuming that the process of mutator fixation occurs on a time-scale much shorter than that associated with environmental changes.  

Our stochastic simulations are based on the well known ``Moran Process'' \citep{moran}.  The following sequence of actions occurs every discrete timestep:
\begin{enumerate}
\item A randomly selected individual is chosen as a potential parent.
\item The chosen individual gives birth with probability proportional to its fitness.  If it does not give birth, the simulation advances to the next timestep.
\item A randomly chosen individual, other than the baby, is killed.  
\item The baby undergoes a deleterious (beneficial) mutation with probability equal to its deleterious (beneficial) mutation rate.  This mutation rate of course depends on whether the baby is a mutator or a wild-type.  Mutations between mutator and wild-type alleles are not allowed.  In effect, this assumes that mutators are generated on a time-scale much longer than that of the entire ``competition experiment''. 
\end{enumerate}

\begin{figure}
\includegraphics[width=6in]{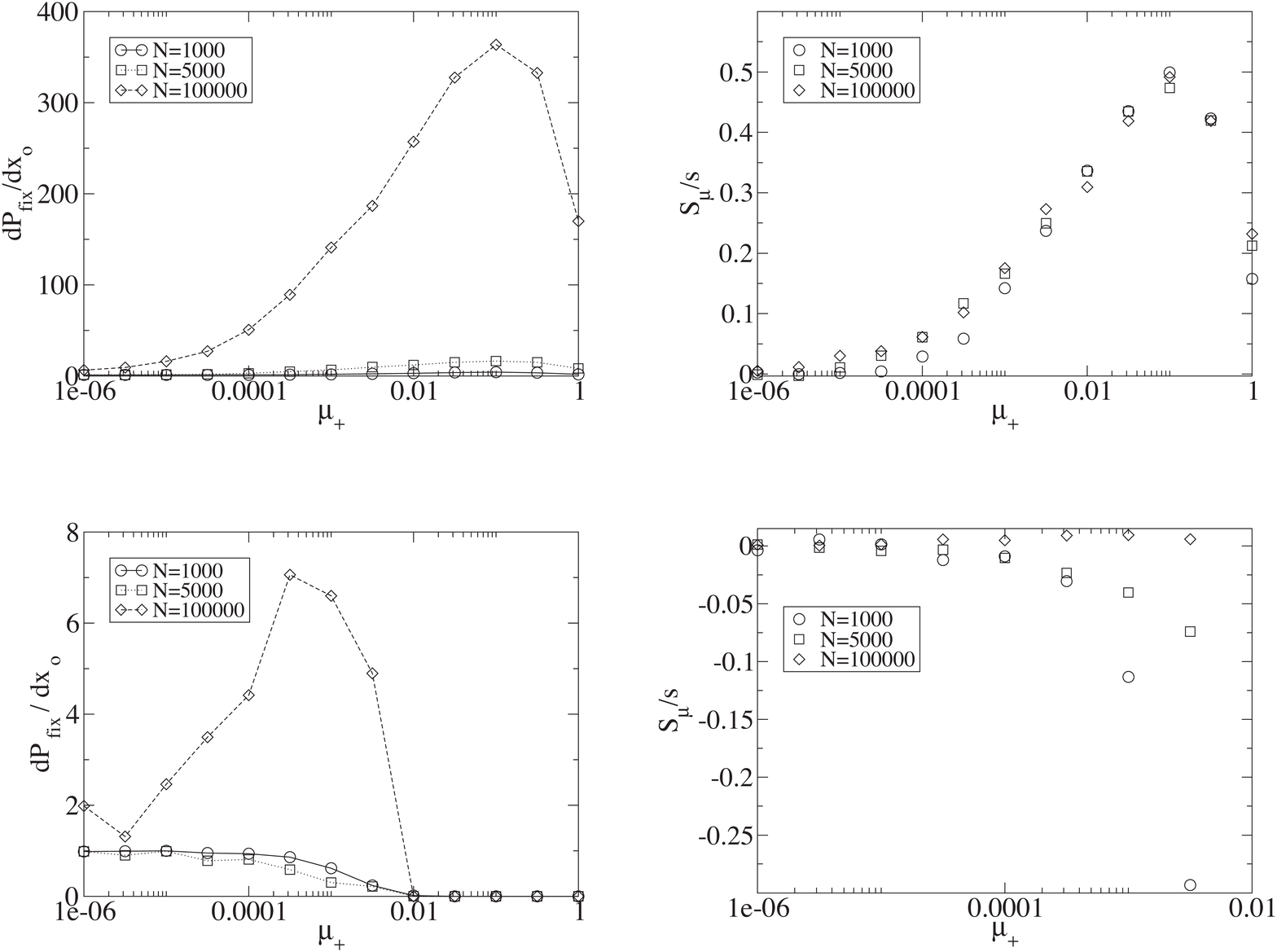}
\caption{\large \doublespacing Averaged results of simulations, and the utility of $S_\mu$ as the measure of mutator success.  When $P_{fix} \ll 1$, $P_{fix}$ increases linearly with $x_o$ (data not shown).  The left panels show the (least squares) slope of said linear increase when the population is well adapted (bottom) and poorly adapted (top) to its environment.  The data on the bottom row are quite noisy because of the small number of trials resulting in fixation.  The panels on the right express the same data, but in terms of the effective selection coefficient $S_\mu$ of the mutator allele obtained by inverting Eq.\ref{classical_fix}.  Whereas the values from the left obviously depend on $N$, the values on the right panels are \textit{independent of N} when $NS_\mu \gg 1$.  This suggests that $S_\mu$, which exposes an underlying simplicity to the simulation results, is a more natural measure of mutator success than $P_{fix}$.  Notice that when the mutator is favored, $S_\mu$ is always less than the selective advantage $s$ of a single beneficial mutation;  this is due both to deleterious mutations and loss due to random drift.  Parameters are $ s=1/120, \mu_-=0, \delta=0, \alpha = .4$ (top) and $.008$ (bottom).  See Supplementary Information for details concerning averaging. }
\label{sim_data}
\end{figure}

\begin{figure}
\includegraphics[width=6in]{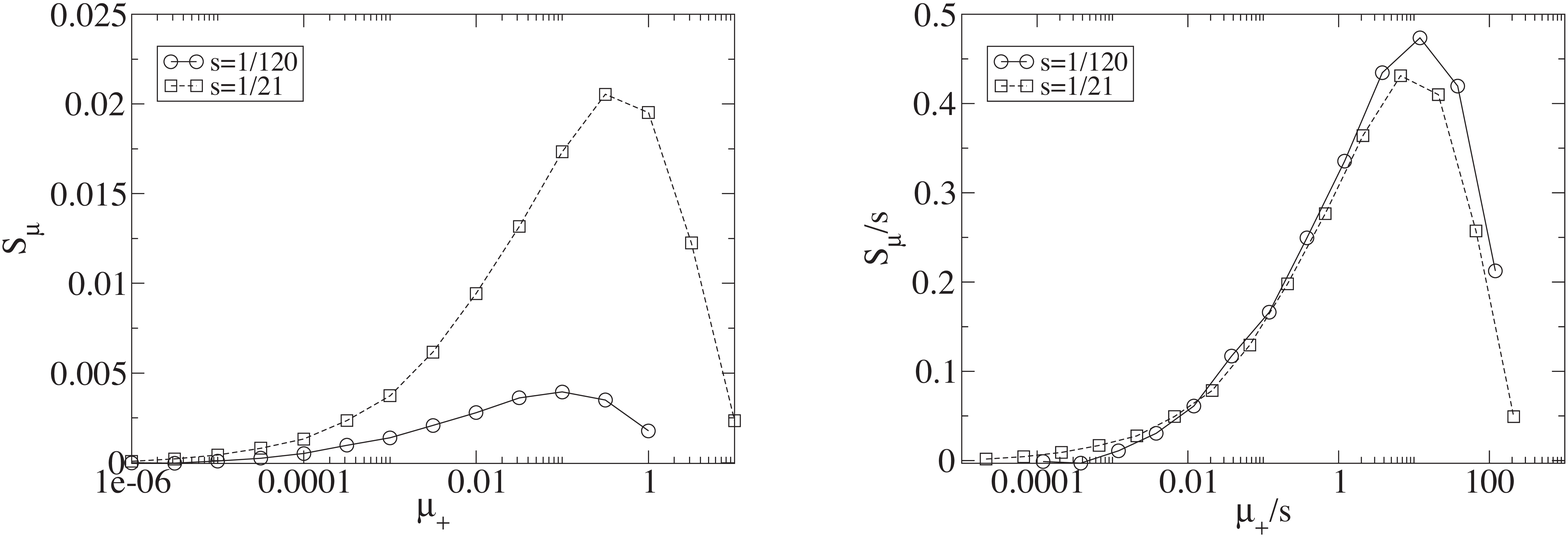}
\caption{\large \doublespacing Dependence on the underlying selective advantage $s$.  The data corresponding to two values of $s$, i.e. two values of $L$, approximately collapse onto a single curve when $S_\mu$ and $\mu_+$ are each scaled by $s$.  The scaling of the independent variable underscores the fact that mutator success for fixed $\alpha$ is largely controlled by the ratio of timescales for mutation ($1/\mu_+$) and selection ($1/s$).  In particular, the sharp decrease in $S_\mu$ at large $\mu_+$ occurs when these timescales become comparable, i.e. when deleterious mutations accumulate in an expanding lineage before it has sufficient time to achieve fixation.  Parameters are $N=5000$, $\mu_- = 0$, $\alpha = .4$, $\delta = 0$.}
\label{s_scale}
\end{figure}

We model the genome of each individual as a string of $L$ bits ~\citep{crosby1970egd, woodcock1996pem, tsimring1996rve}.  A fraction $\delta$ of these bits correspond to critical sites in the genome that, when mutated, cause a lethal phenotype.   In this case, the baby is never born, and the simulation simply advances to the next time-step.  Changing the value of $\delta$ in effect allows for some adjustment of the distribution of deleterious mutational effects.  The birth probability per unit time, which we denote $r$, is proportional to the log-fitness of the chosen individual and equals the fraction of 1's in the genome, denoted by $b/L$.  Key parameters are $\alpha \equiv 1-b/L$ and $\alpha_e \equiv (1-\delta)\alpha$, i.e. the fraction of sites that would be beneficial if mutated.  Thus, all non-lethal mutations have the same strength and genes do not interact.  This scheme for assigning fitness to genotypes is known as a ``multiplicative Fujiyama'' fitness landscape, and is the $K=0$ version of Kaufman's ``NK'' model \citep{kauffman1993oos}.  This toy landscape is obviously a useful mathematical simplification.  Additionally, recent experimental work by \cite{hegreness2006epi,d_fisher_short} shows that some dynamics of real bacteria and yeast populations can be captured by considering mutations of only a single strength.  

Mutation is implemented by ``flipping'' bits with a probability $\frac{\mu_\pm}{L}$ per bit per birth event, depending on whether the baby is a mutator ($+$) or a wild-type ($-$).  The total number of flips is determined by drawing a binomially distributed random number with success probability $\frac{\mu_\pm}{L}$ and number of trials $L$.  Each mutation has a probability $\delta$ of being lethal.  If no mutations are lethal, the number that are beneficial is determined by drawing another binomially distributed random number with success probability $\alpha$ and number of trials equal to the number of flips.  Unless $\mu_\pm$ is  $O(1)$, the probability of more than one mutation occurring during a single birth event is negligible and we will refer to the genome-wide mutation rate as $\mu_\pm$.  

Another useful parameter is $s=\frac{1}{L}/(1-\alpha)$ which, like $\alpha$ and $\alpha_e$, changes throughout the simulation as the population evolves.  We emphasize that this fitness dependent value of $s$ does not represent an epistatic effect.  Rather, it is a consequence of mutations which result in a fixed, additive increment in ``log-fitness.'' 

A consequence of our genomic model is that both the beneficial and deleterious mutation rates will be larger than values encountered in biological populations unless $L$ is extremely large.  While this may seem like an unnecessary and undesirable restriction, it will turn out that our analytic results, which readily handle arbitrary values of the mutation rates, are insensitive to these details of our bit string simulation model.  
\\
\begin{center}{SIMULATION RESULTS}\end{center}

To simplify matters, we first investigate the case where the wild-type mutation rate is zero; results for the more general case will be given later. Fig.\ref{sample_sims} shows typical runs for this case.  These graphs make it clear that if the mutator mutation rate, $\mu_+$, is sufficiently small, the mutator allele hitchhikes to fixation with a single beneficial mutation.  This simple observation reminds us that mutator fixation or loss is not the result of winning the race up the fitness landscape, but rather hitchhiking with beneficial mutations.  Thus, mutator alleles are better thought of as \textit{consequences} of asexual evolution than \textit{causes} of more rapid evolution \citep{sniegowski2000emr}.  When $\mu_+$ is larger, the dynamics are more complex.  Despite this complexity, we will later show, via the success of our analytic approximation scheme, that the fixation process is triggered mostly by the first beneficial mutation to escape random drift.  

\textbf{Dependence on $\mu_+$:} Fig.\ref{sim_data} presents simulation results for three different population sizes and two different degrees of adaptation.  The fundamental measured quantity is the fixation probability $P_{fix}$ of an initially rare mutator.  When $P_{fix} \ll 1$, the mutators are completely independent of one another and $P_{fix}$ increases linearly with $x_o$ (data not shown).  To normalize against the effect of $x_o$, we consider the slope of said linear increase, $dP_{fix}/dx_o$, which equals the mean number of mutator descendants left by each mutator, as our preliminary measure of mutator success.  Fig.\ref{sim_data} (left panel) shows how $dP_{fix}/dx_o$ depends on $\mu_+$.   The small and large $\mu_+$ limits make qualitative sense:  as $\mu_+ \rightarrow 0$, the mutator phenotype is ``turned off'' and therefore neutral, resulting in $dP_{fix}/dx_o \rightarrow 1$.  On the other hand when $\mu_+ \gtrsim 1$, a mutation occurs nearly every birth event and the fitness of an evolutionary line of individuals takes a biased random walk toward the much lower fitness of a completely random genome.  Thus, although it is computationally prohibitive to measure a negligible fixation probability, it is clear that the mutator allele is nearly lethal at sufficiently large $\mu_+$. 

\textbf{Dependence on $N$, and mutator effective selection coefficient:} Fig.\ref{sim_data} also shows that $dP_{fix}/dx_o$ increases with increasing $N$.  This behavior is incompatible with Eq.\ref{freq_dependent}, which is independent of $N$, but is fully consistent with Eq.\ref{classical_fix}:  
\begin{eqnarray}
\nonumber P_{fix} = \frac{1-e^{-Nx_oS}}{1-e^{-NS}}
\end{eqnarray}
We now quantitatively consider whether Eq.\ref{classical_fix}, which applies to mutants with a \textit{direct} fitness advantage, also describes mutators with \textit{indirect} fitness effects.  For this to be the case, the fixation probability measured from simulations with differing values of $N$ and $x_o$ would all correspond to a single value of $S_\mu(\alpha,s,\mu_+,\mu_-,\delta)$.  Using the values of $P_{fix}$ measured from simulations, we used a computer to invert Eq.\ref{classical_fix}, thereby obtaining corresponding values of $S_\mu$.  Fig.\ref{sim_data}(right) shows that, when $NS_\mu \gg 1$, there indeed exists an underlying quantity $S_\mu$, which we call the ``effective mutator selection coefficient,'' that remains invariant as $N$, $x_o$, and $P_{fix}$ change. 

There are several advantages to using $S_\mu$ as the measure of mutator success.  First, it allows Eq.\ref{classical_fix} to determine in advance how $P_{fix}$ depends on $N$ and $x_o$, thereby reducing our number of parameters by two.  Secondly, it allows us to apply aspects of our conceptual understanding of direct mutants to the fixation of indirect mutators.  For example, when $NS_\mu \gg 1$, $P_{fix}$ for a single mutator becomes independent of $N$, i.e. the notion of a frequency independent per capita fixation probability makes sense.  Thirdly, the existence of $S_\mu$, in the sense of Eq.\ref{classical_fix}, invites future questions.  For example, one may wonder whether $S_\mu$, in addition to determining $P_{fix}$, also describes the average \textit{dynamical} behavior of the mutator subpopoulation, e.g. whether $\langle x(t) \rangle \sim e^{S_\mu t}$ when rare.  In this article we do not apply such an interpretation on $S_\mu$.  Rather, we merely interpret it as a succinct descriptor of mutator success.

\textbf{Dependence on strength of mutations:} Fig.\ref{s_scale} shows how $S_\mu$ depends on the strength of the mutations on our fitness landscape, as measured by $s$.  Fig.\ref{s_scale} (left) shows that as $s$ is increased, $S_\mu$ also increases, and reaches its maximum value at a faster mutation rate.  Fig.\ref{s_scale}(right) demonstrates that the curves in the left panel are not as different as they appear: when $S_\mu$ and $\mu_+$ are each scaled by $s$, the curves become nearly identical.  This means that $S_\mu$ is directly proportional to $s$, and that $S_\mu$ is governed by the single composite parameter $\mu_+/s$ rather than $\mu_+$ and $s$ separately.  Thus, an examination of the simulation data has allowed us to reduce our number of parameters by three.  
\\
\begin{center}{INSTANTANEOUS SINGLE LOCUS APPROXIMATION (ISLA)}\end{center}

Stochastic simulations provide valuable signposts along the way to understanding mutator fixation.  However, a deeper understanding, as well as the ability to probe computationally prohibitive regions of parameter space, requires an analytic approach as well.  At a given time, the state of the population is fully specified by (i) the number of mutators, (ii) the fitness distribution of the wild-type subpopulation, and (iii) the fitness distribution of the mutator subpopulation.  A complete solution to the stochastic process requires an enumeration of the transition probabilities between each of these states at each point in time.  The problem with such an approach is the extremely large number of possible fitness distributions and the correspondingly high dimensionality of the resulting governing differential equations.  In order to make progress, we note the heuristic rule that deleterious mutations are rapidly removed from the population, whereas beneficial mutations, and all loci linked to them, become rapidly fixed.  This observation motivates the following approximations that handle mutations, which are the ultimate source of the aforementioned daunting multiplicity of fitness distributions.  

\noindent
\textit{Approximation 1 (A1)}: We assume that when a beneficial mutation arises, it instantly becomes fixed with a probability given by the classical fixation probability $\pi$ of a beneficial mutation in a static, homogeneous environment.  For our Moran process dynamics, this probability is simply $s$ if $s\ll1$ and $N s \gg 1$.  All loci in the genome in which the beneficial mutation arose also achieve fixation via hitchhiking.  This represents the most common process by which the mutator allele achieves fixation or loss.  

\noindent
\textit{Approximation 2 (A2)}: The remaining fraction $1-s$ of beneficial mutations are simply ignored and treated as if no mutation occurred.  This approximation is necessarily somewhat awkward.  On the one hand, A2 is unnatural in that it allows lineages which are destined to be extinguished by random drift to remain in the population and potentially generate their own beneficial mutants.  An alternative, which we call A2$^*$, is to immediately kill the beneficial mutants which do not sweep, which is clearly too harsh.  These two alternatives lead to a trivial difference in our formulas, and are discussed in Supplementary Material.

\noindent
\textit{Approximation 3 (A3)}: Deleterious mutations are treated as effectively lethal, since their descendants are quickly removed from the population. This results in an effective reduction in the birthrate of the mutator strain.

Since these approximations preclude fitness polymorphism over finite time intervals, they allow us to describe the dynamics of the entire population with the single time dependent random variable $x$, i.e. the frequency of the mutator locus.  Approximating $x$ as a continuous variable, and expressing time in ``generations,'' the diffusion equation governing $P(x,t)$ is (see Supplementary Information for a detailed derivation)

\begin{eqnarray}
\nonumber \frac{\partial P}{\partial t} &=&\frac{1}{N}\frac{\partial^2}{\partial x^2}\left[x(1-x)P\right]\\
\nonumber &+& (\mu_+-\mu_-)\left[1-\alpha_e(1-s)\right]\frac{\partial}{\partial x}\left[x(1-x)P\right] \\
&-& N\alpha_e s\left[x\mu_++(1-x)\mu_-\right]P
\label{full_fpe}
\end{eqnarray}  
Each of the three lines in Eq.\ref{full_fpe} has a straightforward physical interpretation.  The first line represents ``random genetic drift.''  The second line represents the mutational load of the mutator.  The final line represents the ``decay'' of probability from the open interval $x\in(0,1)$ due to beneficial mutations that instantaneously sweep.  

An approximation to a limited version of Eq.\ref{full_fpe} is solved in Supplementary Information.  However, we can write an equivalent ``backward Kolmogorov'' equation which is often more mathematically convenient than Eq.  \ref{full_fpe}.   Defining $G(x_o,t)$ as the probability that the mutator has been \textit{lost} by time $t$, given that $x=x_o$ at $t=0$, we find 
\begin{eqnarray}
\nonumber \frac{\partial G(x_o,t)}{\partial t} &=& \frac{1}{N}x_o(1-x_o)\frac{\partial^2}{\partial x_o^2}G(x_o,t) \\
 \nonumber  &-&(\mu_+-\mu_-)\left[1-\alpha_e(1-s)\right]x_o(1-x_o)\frac{\partial}{\partial x_o}G(x_o,t)\\
    &-& N\mu_+\alpha_e s x_o G(x_o,t) +N\mu_-\alpha_e s (1-x_o) (1-G(x_o,t))
\label{tdep_back_eq}
\end{eqnarray}
The backward equation is primarily useful in its steady state form.  Defining $G(x_o,t\rightarrow \infty) \equiv G_\infty(x_o)$ and taking the continuum limit, we obtain the ODE
\begin{eqnarray}
 \nonumber0 &=& \frac{1}{N}\frac{d^2}{dx_o^2}G_\infty \\
 \nonumber  &-&(\mu_+-\mu_-)\left[1-\alpha_e(1-s)\right]\frac{d}{dx_o}G_\infty\\
    &-& N\mu_+\alpha_e s\frac{G_\infty}{1-x_o} +N\mu_-\alpha_e s \frac{1-G_\infty}{x_o}
\label{back_eq}
\end{eqnarray}

\begin{figure}
\includegraphics[width=6in]{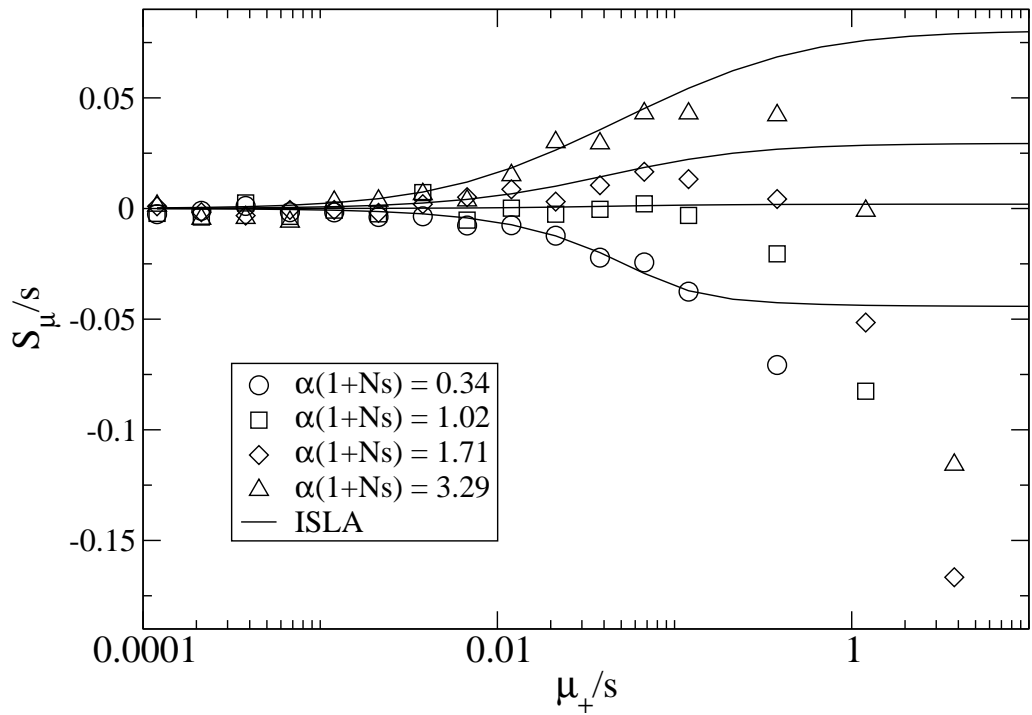}
\caption{\large \doublespacing Behavior near the transition from favored to disfavored mutators.  When $\alpha_e$ is greater than a critical value $\alpha_e^{crit}$, the mutator allele is favored ($S_\mu > 0$) for small enough $\mu_+$.  Our analytic approach (ISLA) predicts that the transition occurs at $(Ns+1)\alpha_e^{crit} = 1$, which agrees extremely well with simulation data.  Parameters are $N=5000,s=1/120,\mu_-=0,\delta=0$.  The number of available beneficial mutations are, in order of decreasing mutator success: 10, 5, 3, and 1.} 
\label{trans_fig}
\end{figure}
\begin{figure}
\includegraphics[width=6in]{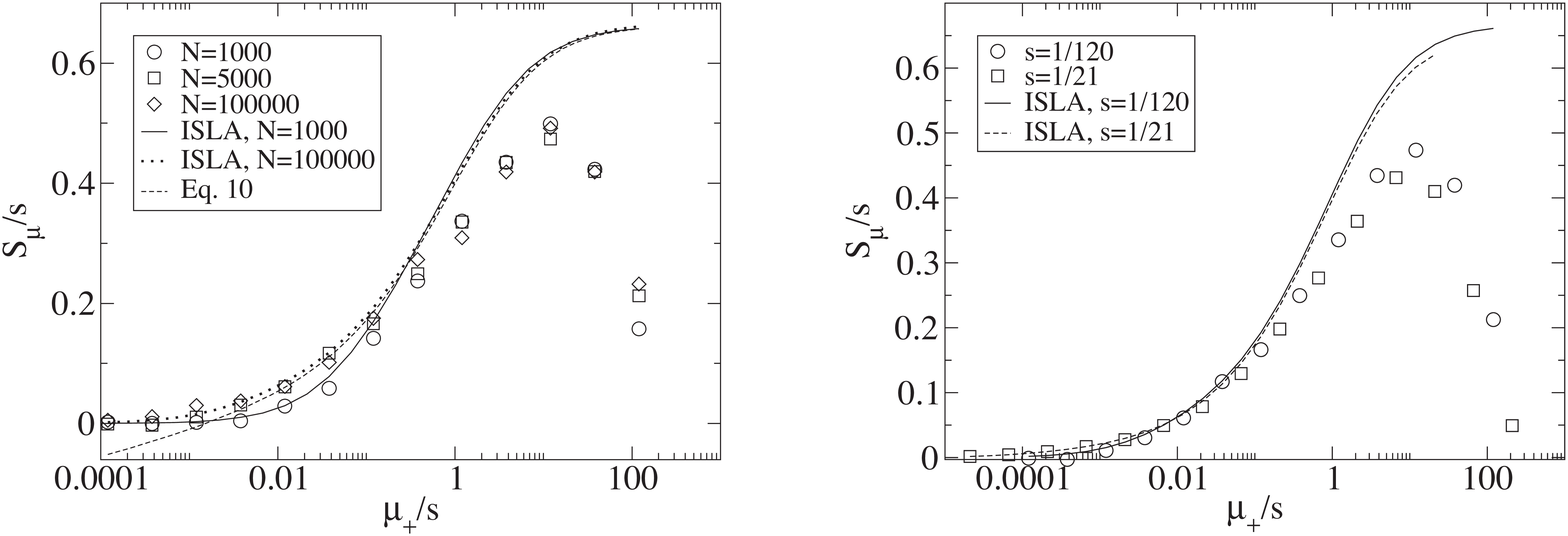}
\caption{\large \doublespacing Comparison of simulation, numerical solution of Eq.\ref{back_eq}, and the analytic approximation Eq.\ref{appx_S_mu}.  The exact numeric solutions to our ISLA Eq.\ref{back_eq} for different $N$ converge to the analytic approximation Eq.\ref{appx_S_mu} when $NS_\mu \gg 1$ (left).   Solutions to Eq.\ref{back_eq} show, in agreement with simulation, that $S_\mu/s$ depends on $\mu_+/s$ rather than $\mu_+$ and $s$ seperately (right).  Parameters are those used in Figs. \ref{sim_data}, \ref{s_scale}.} 
\label{mod_sim_comp_fig}
\end{figure}

\textbf{Solution and Analysis Without Wild-Type Mutations}: We return for now to the simpler case $\mu_- =0$, deferring until later the more general situation. Eq.\ref{back_eq} can be solved exactly in terms of the Whittaker $M$ function\citep{abromsteg}.  This exact solution is however not immediately instructive (and in any case cannot be generalized to the case of finite wild-type mutation rate). It is simpler in practice to solve Eq.\ref{back_eq} numerically (see Supplementary Information).  It is also possible to extract some useful information directly from the differential equation.

First, we note that a simple analysis  reveals when the mutator allele will be favored.  For notational convenience we define the constants
\begin{eqnarray}
\nonumber B &\equiv& \mu_+\left[1-\alpha_e(1-s)\right]\\
\nonumber C &\equiv& \mu_+\alpha_e s
\end{eqnarray}
According to ISLA, the mutator is neutral \textit{for all $\mu_+$} when $G_\infty(x_o) = 1-x_o$.  Plugging this into Eq.\ref{back_eq}, we find that this requires $B=NC$, or
\begin{equation}
\alpha_e^{crit}=\frac{1}{1 + (N-1)s} \approx \frac{1}{1+Ns}
\label{transition}
\end{equation}
It is easy to check that this expression also holds for the $\mu_- > 0$ case.  First note that Eq.\ref{transition} is equivalent to our heuristic guess, Eq.\ref{game_theory_trans}, if $Ns \gg 1$.   Examining Eq.\ref{transition}, we see that conditions which favor the emergence of mutators (at least when the resident mutation rate $\mu_-$ is negligibly small) are large population size, potent mutations, and a relatively large fraction $\alpha_e$ of sites that would be beneficial if mutated, perhaps due to an environment to which the organism is not well adapted. The fact that large $\alpha_e$ favors mutators is obvious.  The dependence on $N$ is simply a result of the fact that as population size increases, the neutral fixation probability $1/N$ becomes an easier benchmark to exceed.  The qualitative dependence on $s$ is also straightforward in hindsight, given A1-A3: increasing $s$ increases the fraction of beneficial mutations that achieve fixation, but does not affect the fate of deleterious mutations, all of which are treated as lethal.  Also notice that for sufficiently large $N$ the mutator is always favored, although its fixation probability may be very small: it is favored only in the sense that it fares better than a neutral allele whose fixation probability is $1/N$.  Fig.\ref{trans_fig} demonstrates the success of Eq.\ref{transition} when $\mu_+/s \ll 1$.  The failure of ISLA for larger $\mu_+/s$ will be discussed later.  We next develop approximate solutions to Eq.\ref{back_eq}, with $\mu_-=0$.

\textit{Strongly Favored Mutators ($NS_\mu \gg 1$)}:  In this regime, we expect $P_{fix}$ to increase rapidly with $x_o$.  Therefore, we expect the loss  probability $G_\infty(x_o)$ to decrease rapidly, and $1/(1-x_o)$ to differ significantly from $1$ only when $G_\infty \approx 0$.  Then, for $x_o \ll 1$, we can approximately take  $1-x_o \rightarrow 1$, and the solution to Eq.\ref{back_eq} with $\mu_- = 0$ is simply
\begin{eqnarray}
G_\infty(x_o) &=&e^{-Nzx_o} \label {expo_loss_appx}\\
\nonumber z &\equiv&\frac{\sqrt{B^2 +4C} -B}{2} 
\end{eqnarray}
Our approximation is self consistent if indeed $G_\infty$ decays rapidly, i.e.  $Nz \gg 1$.  
This solution does not satisfy the boundary condition at $x_o = 1$ since our solution is only valid for $x_o\ll 1$.  Beyond this region the structure of the solution is more complicated, which need not concern us here since fixation is essentially total in this regime. We then have for the fixation probability of the mutator
\begin{equation}
P_{fix}(x_o) = 1-e^{-Nzx_o} \quad\quad\quad (Nz\gg 1)\label{expo_fix_appx}
\end{equation}
A comparison with Eq.\ref{classical_fix} shows that, according to A1-A3 and in the limit $Nz \gg 1$, the mutator effectively behaves like a simple advantageous mutant with a well defined selection coefficient $S_\mu = z$:
\begin{equation}
S_\mu  = z =\frac{\sqrt{B^2 + 4C}-B}{2} \approx \frac{\mu_+}{2}\left[\sqrt{(1-\alpha_e)^2 + 4\alpha_e s/\mu_+}-(1-\alpha_e)\right]\quad\quad\quad NS_\mu \gg 1
\label{appx_S_mu}
\end{equation}
A comparison of the stochastic simulation data with both a numerical solution of Eq.\ref{back_eq} and this approximate analytic expression (Eq.\ref{appx_S_mu}) is given in Fig.\ref{mod_sim_comp_fig}.
We see that our approximate $S_\mu/s$ only depends on $\mu_+/s$ rather than $\mu$ and $s$ separately, as we noted in the Simulation Results section.  

\begin{figure}
\includegraphics[width=6in]{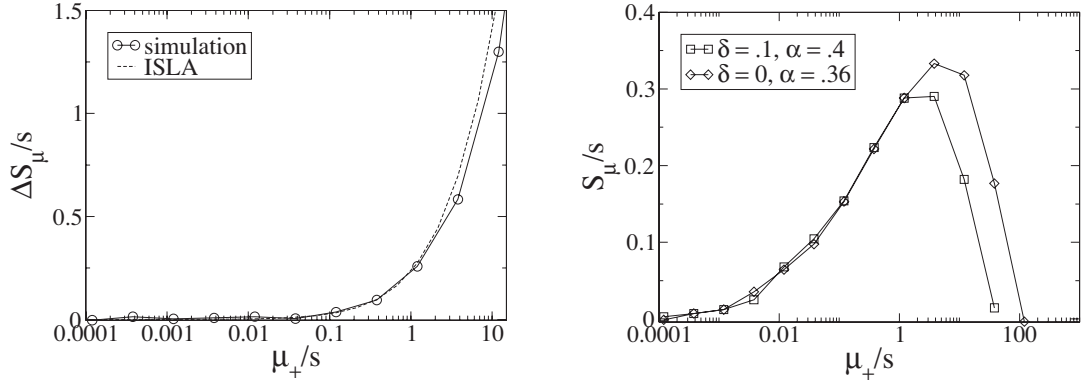}
\caption{\large  \doublespacing The role of non-lethal deleterious mutations.  We ``turned off'' deleterious mutations, both in simulations and in ISLA, by setting the deleterious mutation rate to zero and leaving the beneficial mutation rate unchanged (left).  The difference between these results and the corresponding ones \textit{with} deleterious mutations is plotted on the vertical axis on the left.  For $\mu_+/s \lesssim 1$, deleterious mutations have the same effects in ISLA Eq.\ref{back_eq} as in simulations (left).  ISLA essentially treats deleterious mutations as lethal (A3), instead of merely having a selective disadvantage $-s$.  We tested this approximation directly in simulations by varying the parameters $\alpha$ and $\delta$ while holding the product $\alpha(1-\delta) \equiv \alpha_e $ constant (right).  Parameters are $s=1/120,N=5000, \mu_-=0$ and $ \alpha = .4,\delta=0$ (left only).}
\label{lethal_comp}
\end{figure}

For small  $\mu_+ \ll \alpha_e s$, $C\gg B^2$ and $S_\mu\approx \sqrt{C} = \sqrt{\mu_+ \alpha_e s} $, and thus only advantageous mutations are relevant to mutator success.  This result (which is directly supported by Fig.\ref{lethal_comp} to be discussed later), shows that in this regime, random drift, and not deleterious mutations, is the only check on mutator success.

In the complementary regime where $\mu_+ \gg \alpha_e s, |S_\mu|$ approaches its maximum value $S^*_\mu$ with respect to $\mu_+$.  Here, the solution is the same as if the second derivative term, which represents random drift, were dropped from Eq.\ref{back_eq} (see below).  Therefore, random drift is irrelevant in this regime and deleterious mutations alone limit mutator success, giving 
\begin{equation}
S^*_\mu = \frac{C}{B} \approx \frac{\alpha_e }{1-\alpha_e}s
\label{s_max}
\end{equation}
The factor in Eq.\ref{s_max} multiplying $s$ is the ratio of beneficial mutations to deleterious and lethal mutations.  In real biological populations, this ratio is certainly less than one, and hence $S^*_\mu \ll s$.

\textit{Marginal Mutators ($NS_\mu  \lesssim 1$)}: We can readily make progress in this regime if $N\mu_+  \gg 1$ and $N^2 \mu_+ \alpha_e s \gg 1$.  In this case, the $B$ and $C$ terms dominate Eq.\ref{back_eq} and the solution for $G_\infty$ is simply
\begin{equation}
G_\infty(x_o) \approx (1-x_o)^{NS^*_\mu}
\label{large_mu_thresh}
\end{equation}
with a fixation probability $P_{fix}(x_o)\approx Nx_oS_\mu^*$.  In obtaining this solution, we dropped the second derivative term in Eq.\ref{back_eq}, which could in principle introduce large errors near $x_o=1$, where $G''(x_o)$ from Eq.\ref{large_mu_thresh} is in fact large.  Nonetheless, it turns out that Eq.\ref{large_mu_thresh} satisfies the boundary condition at $x_o = 1$ and thus remains a valid leading order approximation for all $x_o$.  Since $P_{fix}$ is comparable to $1/N$ in the present marginal case, we cannot interpret $S_\mu^*$ as a mutator selection coefficient here.  Rather, we have $P_{fix} = x_o(1+NS_\mu/2)$, from which we obtain $NS_\mu = 2\frac{\alpha_e(Ns+1)-1}{1-\alpha_e}$, independent of $\mu_+$. The numerator of this expression makes clear the agreement with our previous estimate for the critical value of $\alpha_e$ given by Eq.\ref{transition}.

The case where $N\mu_+ \lesssim 1$ and $NS_\mu \lesssim 1$ requires a more lengthy analysis, and is presented in Supplementary Information.

\begin{figure}
\includegraphics[width=6in]{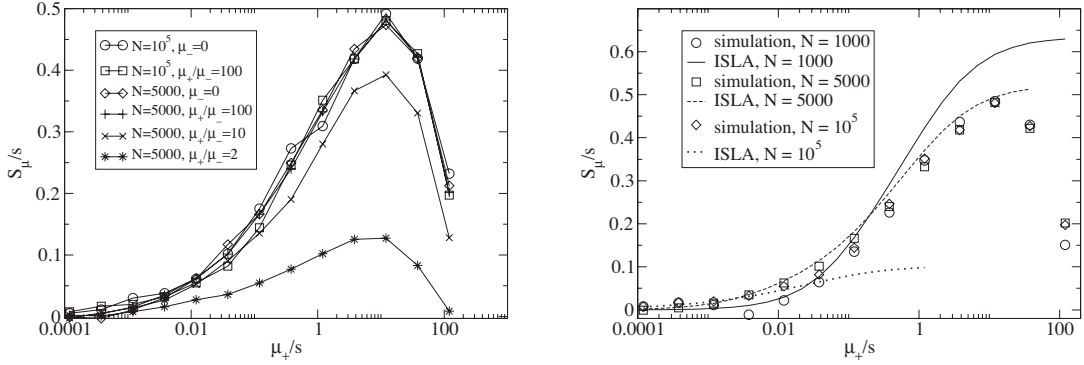}
\caption{\large  \doublespacing Small effect of mutations arising in wild-type backgrounds.  ISLA predicts that these mutations will become important in the weak-effect mutator regime defined by $\frac{R(1-\alpha_e)}{N\alpha_e s}\lesssim 1$, where $R \equiv \mu_+/\mu_-$.  However, the simulation data show that  mutations in wild-type backgrounds sometimes have a negligible impact even in the weak-effect mutator regime.  In the panel on the right, $\frac{R(1-\alpha_e)}{N\alpha_e s}$ has the values 18, 3.6, and .18, respectively, as $N$ is increased.  Accordingly, ISLA predicts a decrease in $S_\mu$, but $S_\mu$ did not change in simulations.  The panel on the left shows that beneficial mutations in wild-type backgrounds eventually decrease $S_\mu$ for large enough $R$, though the decrease here is smaller than what ISLA predicts.  Parameters are $\alpha = .4, s=1/120, \delta=0$, and $\mu_+/\mu_- = 100$ (right). 
 } 
\label{mum_comp_fig}
\end{figure}

\begin{center} EFFECT OF WILD-TYPE MUTATIONS \end{center}

We now turn our attention to the more complicated case when mutations in wild-type backgrounds are allowed, i.e. $\mu_- > 0$.  We begin by solving Eq.\ref{back_eq} for $\mu_- > 0$ in the large $N\mu_{\pm}$ limit, where the second derivative term can be neglected.  Working in this limit simplifies the mathematics, and is sufficient for illustrating the points that we intend to make.  An approximation that incorporates the second derivative term and random drift is included in Supplementary Information.  In the large $N\mu_{\pm}$ limit, 
\begin{eqnarray}
\nonumber0 =&-& (\mu_+-\mu_-)\left[1-\alpha_e(1-s)\right]\frac{d}{dx_o}G_\infty\\
\nonumber  &-& N\mu_+\alpha_e s\frac{G_\infty}{1-x_o} \\
\nonumber  &+& N\mu_-\alpha_e s \frac{1-G_\infty}{x_o} 
\end{eqnarray}
This first order, linear ODE can be solved by standard methods.  Defining $R \equiv \mu_+/\mu_-$, we obtain 
\begin{eqnarray}
P_{fix} \approx Nx_o s\frac{\alpha_e}{1-\alpha_e} \left(
                                      1 +  \frac{\alpha_e(N s + 1)-1}{R(1-\alpha_e)}\right)^{-1}+O(x_o^2)
\label{asymp_mum}
\end{eqnarray}

The prefactor in Eq.\ref{asymp_mum} is identical to our previous expression for the $\mu_-=0$ case (Eqs.\ref{s_max},\ref{large_mu_thresh}) when $x_o \ll 1$.  Recall that the sign of the quantity $\alpha_e(N s + 1)-1 \approx N \alpha_e s - 1$ determines whether mutators are favored (Eq.\ref{transition}).  Therefore, mutations in wild-type backgrounds decrease $P_{fix}$ when mutators are favored and \textit{increase} $P_{fix}$ when they are disfavored.  This latter effect occurs because mutating is generally a losing strategy when $\alpha_e(Ns + 1)-1 < 0$ (see Eq.\ref{game_theory_trans}): the small persistent cost of deleterious mutations exceeds the huge occasional benefit of a selective sweep.  Thus, in this regime the wild-type aids the mutator by participating in this losing strategy. 
 
Eq.\ref{asymp_mum} also determines when $R$ is sufficiently large to ignore mutations in wild-type backgrounds.  In other words, Eq.\ref{asymp_mum} allows us to define natural ``strong-effect'' and ``weak-effect'' mutator regimes.  For weak-effect mutators, $\frac{\alpha_e(N s + 1)-1}{(1-\alpha_e)} \approx N\alpha_e s \gg R$, and Eq.\ref{asymp_mum} reduces to $P_{fix} = x_oR$, which is \textit{independent of $N$}.  This is the same as Eq.\ref{freq_dependent} for $x_o \ll 1$.  Thus, in this regime, ISLA predicts that mutational competition with the wild-type is the dominant factor limiting mutator fixation, and we recover the explicitly frequency dependent heuristic picture.  In the opposite extreme of strong-effect mutators, regardless of the sign of $\alpha_e(N s + 1)-1$, we recover our $\mu_-=0$ result (Eqs.\ref{s_max},\ref{large_mu_thresh}) where deleterious mutations are the dominant factor limiting mutator fixation.

These are pleasing mathematical results that seem to reconcile opposing heuristic viewpoints.  However, they do not always match simulations in the weak-effect mutator regime.  Fig.\ref{mum_comp_fig} (right) shows numerically generated solutions to Eq.\ref{back_eq} (Eq.\ref{asymp_mum} gives the large $\mu$ limit of these curves) as compared to the outcome of simulations.  The disagreement is obvious: ISLA drastically overestimates the effect of the mutations in wild-type backgrounds.  Fig.\ref{mum_comp_fig} shows that beneficial mutations in wild-type backgrounds eventually decrease $S_\mu$ for large enough $R$, though the decrease here is smaller than what ISLA predicts.  The small effect of these mutations persisted even when we used parameters such that the wild-type subpopulation generated mutations at a rate $N(1-x_o)\mu_-$ that was equal to or even greater than the corresponding rate $Nx_o\mu_+$ in the mutator subpopulation.  Although we do not fully understand this discrepancy, we can point to its source: There is a subtle error involving the final term of both Eqs.\ref{full_fpe},\ref{tdep_back_eq} which states that during a single time-step, the mutator has a probability $(1-x)\mu_- \alpha_e s$ of becoming \textit{instantly} lost.  This is incorrect.  The correct statement is that $(1-x)\mu_- \alpha_e s$ is the probability that during one time-step, the wild-type generates a beneficial mutation that will \textit{eventually} escape loss to random drift.  Such mutations sweep through the population during a mean time interval $t_{sweep} \sim \frac{\ln(Ns)}{s}$ generations which is typically much longer than the time to extinction of a mutator due to random drift $\bar{t}_{drift} \approx \ln(N)$ \citep{crow:ipg}.  However, for sufficiently large $s$, $t_{sweep}$ is small, A1 becomes a better approximation, and ISLA more closely matches simulations.  An example of this agreement is presented in Supplementary Information, where $s=1/3, N=1000,\alpha_e=.4,R=10$.  Thus, ISLA provides accurate results except in the weak-effect mutator regime with sufficiently small $s$. Unfortunately, we do not have a quantitative sense as to how large $s$ must be in order to achieve accuracy.  We plan to address this issue in future work.  



In Supplementary Information, we more closely examine the role of $\mu_-$ by presenting and interpreting the distribution of fixation and loss times for mutators when $\mu_-=0$ and $\mu_+/\mu_-=100$.
\\
 
\begin{center} {COMPARISON OF ISLA TO SIMULATION} \end{center}
\label{validity_of_SMA}

We now return to the case $\mu_- = 0$, where the results of ISLA agree with simulations when $R \equiv \mu_+/\mu_-$ is sufficiently large.  Figs.\ref{trans_fig},\ref{mod_sim_comp_fig} illustrate the agreement between ISLA Eq.\ref{back_eq} and simulations, whenever $\mu_+/s$ is not too large.  However, for larger $\mu_+/s$, we see the emergence of two qualitatively distinct discrepancies between ISLA and simulations.  For $\mu_+/s \lesssim 1$, a relatively small difference accumulates, whereas when $\mu_+/s$ reaches values of $O(1)$, a drastic difference emerges.  In this section, we analyze the sources of these discrepancies.  

The broad reason that ISLA and simulation do not agree for all $\mu_+$ is simply that A1-A3 and the resulting transition probabilities are only an approximation of the complex stochastic process executed by the simulations.  Indeed, strictly speaking, the simulation does not even undergo a Markov process with respect to the variables $x,t$: one must also consider the fitness distributions of the subpopulations in order to write down the exact transition probabilities.  When viewed this way, it is perhaps surprising that A1-A3 work as well as they do.  We now specifically point out the errors introduced as a result of A1-A3, all of which are associated with mutational processes.

\textit{A3 is accurate when $\mu_+/s \lesssim 1$}: We first analyze the way that ISLA treats deleterious mutations, which includes both A3 (which treats all deleterious mutations as lethal) and A1 (which does not allow deleterious mutations to arise in the course of fixation of an ``evolved'' clone).   Fig.\ref{lethal_comp} (right) compares simulation results from two sets of parameters with identical beneficial mutation rates ($\alpha_e\mu_+$) but different allocations of lethal and deleterious mutations via a difference in the parameter $\delta$.  The results are essentially identical as long as $\mu_+/s \lesssim 1$.  This shows that as far as mutator fixation is concerned,  mutations of effect $-s$ can be considered lethal, i.e.\ A3 is accurate in this regime.  

\textit{A1 is accurate when $\mu_+/s \lesssim 1$}: Furthermore, we can test all the effects of deleterious mutations by removing them from both the simulations and ISLA: the deleterious mutation rate is set to zero whereas the advantageous mutation rate is left unchanged. The results of this case are presented in Fig.\ref{lethal_comp} (left).  Predictably, $S_\mu$ increases monotonically with $\mu_+$ in this case (data not shown).  To compare the effect of deleterious mutations in simulations against those same effects according to ISLA, Eq.\ref{back_eq}, we plot the difference $\Delta S_{\mu} \equiv S_{\mu, no-deleterious} - S_{\mu,deleterious}$ between results with deleterious mutations ``off'' and those with deleterious mutations ``on'' in the two cases.  We see in Fig.\ref{lethal_comp} (left) that $\Delta S_\mu$ from ISLA matches that from simulation until $\mu_+/s \rightarrow 1$.  Also note that $\Delta S_\mu \approx 0$ for $\mu_+/s \ll .1$, illustrating the negligible effect of deleterious mutations in this regime.  Thus, both A1 and A3 are accurate when $\mu_+/s \lesssim 1$.

\textit{A2 fails when $\mu_+/s \lesssim 1$}:  Since A1 and A3 remain valid in this regime, the mild discrepancy between simulations and ISLA must originate in A2, which handles beneficial mutations.  Specifically, the fraction $(1-s)$ of advantageous mutants that are lost to random drift are treated as neutral mutators which can later give rise to beneficial mutants that may sweep through the population.  In some sense, this overstates the potential of these mutants because, in fact, they are typically lost to random drift within a few generations \citep{crow:ipg}.  There is no simple remedy for this deficiency in A2, but an alternative, which we denote A$2^*$, is to immediately kill these advantageous mutants, thereby treating them equivalently to deleterious and lethal mutants.  Whereas A2 overestimates $S_\mu$ in this regime, A2$^*$ underestimates it.  Thus, the simulation data is bounded by the predictions of A2 and A2$^*$ when $\mu_+/s \ll 1$.  See Supplementary Information for a graphical comparison and further discussion of A2$^*$.

\textit{A1 fails when $\mu_+/s \sim 1$:} We now turn to the large discrepancy between ISLA and simulations when $\mu_+/s$ is $O(1)$, as seen in Fig.\ref{lethal_comp}.  Roughly speaking, this occurs when the time-scales of (deleterious) mutation and selection become comparable.  In this regime, members of an expanding ``evolved'' clone are ``lost'' due to deleterious mutations faster than they are ``added'' due to selection.  Consequently, the fixation probability of an advantageous mutant in a homogeneous genetic background $\pi(s) < s$ and A1 fails.  Semi-quantitatively, we expect this effect to set in when $(1-\alpha_e)\mu_+/s \sim 1$.  The $\alpha_e$ dependence can be seen by comparing Figs.\ref{trans_fig},\ref{mod_sim_comp_fig}.
\\
\begin{center}{COMPARISON TO EXPERIMENT}\end{center}

As mentioned previously, the spontaneous emergence of mutator alleles has been documented in laboratory evolution experiments with \textit{E.coli} \citep{sniegowski1997ehm,shaver2002fea}.  In this experiment, mutator alleles with $R \approx 100$ became fixed in 3 out of 12 independently evolving \textit{E.coli} populations within 10,000 generations.  The total number of mutators generated among 12 lines during 10,000 generations is approximately $N_e \times U  \times (10^4 \times 12)$, where $U$ is the mutation rate into the mutator state and $N_e$ is the effective population size \citep{wahl55pbm,wahl2002eip}.  $U$ has been measured between $5 \times 10^{-7}$ \citep{taddei1997mop} and $5 \times 10^{-6}$ \citep{bo_mutator}, and we find $N_e = 6.3\times 10^{7}$ (see Supplementary Information).  Since three of these mutators achieved fixation, the experimental fixation probability $P_{fix,expt}$ is approximately given by $3/(N_e \times U  \times 10^4 \times 12)$ and bounded by 
\begin {equation}
7.9 \times 10^{-8} < P_{fix,expt}  < 7.9 \times 10^{-7} 
\label{expt_pfix}
\end{equation}
This value is 5-50 times that of a neutral allele ($1/N_e$).

In order to compare this value to the predictions of ISLA, we need experimental values for the parameters $\mu_+,\alpha_e,$ and $s$.  It turns out that the equivalent set of parameters $s$, the beneficial mutation rate $\mu_{ben,+} = \alpha_e \mu_+$, and the deleterious mutation rate $\mu_{del,+}=(1-\alpha_e) \mu_+$ are more readily available in the literature.  A survey of these parameter values, as well as a more careful discussion of their meaning, can be found in Supplementary Information.  Presently, we use the beneficial mutation rate $\mu_{ben} = 2.8 \times 10^{-8}$ and selection coefficient $s = .1$ obtained by \cite{lenski1991lte}.  Following \cite{keightley1999tma}, we take $\mu_{del} = 1.6 \times 10^{-1}$.  These mutation rates are based on the measured wild-type values and assume $R=100$.  Since $N_e \mu_{del,+} \gg 1$, $N_e^2 \mu_{ben,+} s \gg 1$, and $N_e \alpha_e s \ll R$, these populations are in the drift-less, strong-effect mutator regime.  Therefore, the appropriate formula  is either Eq.\ref{large_mu_thresh}  or Eq.\ref{asymp_mum}, which give the same results.  Plugging our parameter values into ISLA, we obtain
\begin{equation}
{P}_{fix,isla} = 1.8\times 10^{-8}
\label{best_guess}
\end{equation}
in reasonable agreement with the rough experimental value (Eq.\ref{expt_pfix}).  Other choices for parameter values, particularly $\mu_{ben,+}$, would result in less impressive agreement with experiment.  See Supplementary Information for further discussion.  

It is also interesting to note that, according to these experimental parameters, $N \alpha_e s \approx 1.1$, indicating that these \textit{E. coli} populations only very marginally favored mutators.  This could explain why no mutators fixed during the next $25,000$ generations: $N\alpha_e s$ had decreased below the threshold value of one as fewer, and less potent, beneficial mutations became available.  

Due to the relatively large population size $N_e = 6.3 \times 10^7$ and the anticipated small fixation probability, we cannot obtain an accurate measurement of $P_{fix}$ using our simulation method.  However, for these experimental parameters, $\mu_+(1-\alpha_e)/s = \mu_{del,+}/s$ is $O(1)$ 
and therefore we expect the data to lie in the decreasing portion of curves such as Fig.\ref{mod_sim_comp_fig}.  Thus, our ISLA estimate of $P_{fix}$ is probably much larger than what simulations would yield.  We briefly return to this issue in the Discussion.  
\\

\begin{center}{DISCUSSION}\end{center}
\begin{figure}
\includegraphics[width=6in]{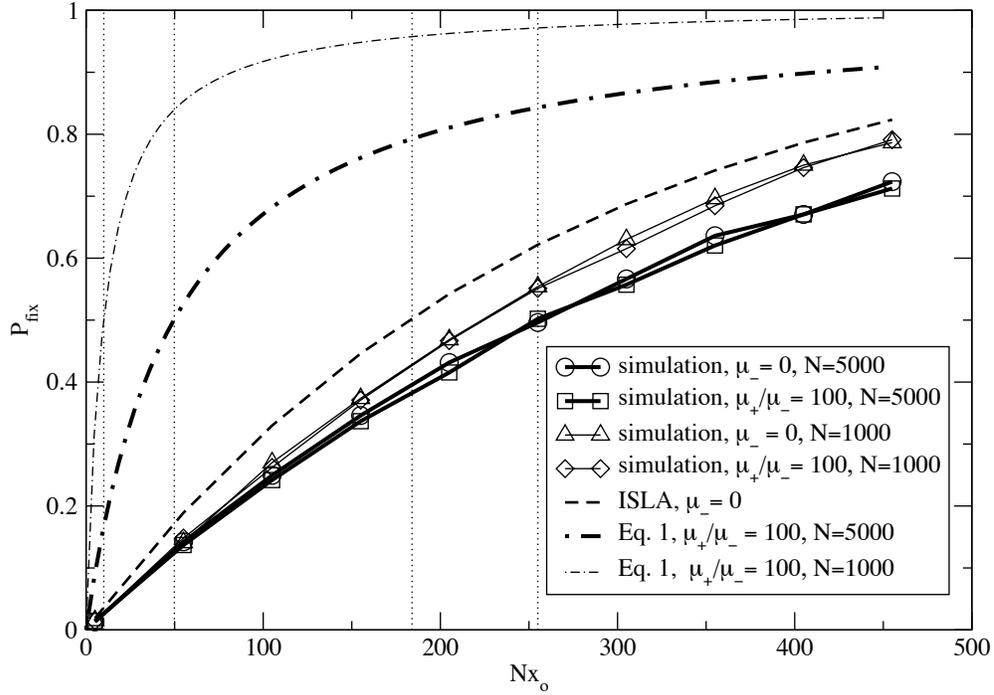}
\caption{\large \doublespacing The scaling behavior of Eq.\ref{freq_dependent} and ISLA are \textit{qualitatively} different.  If the initial number of mutators $Nx_o$ is kept constant while $N$ is increased, then ISLA predicts that $P_{fix}$ remains invariant, whereas the frequency dependent Eq.\ref{freq_dependent} predicts a large change.  Simulations are in better accord with ISLA than Eq.\ref{freq_dependent}.  These scaling predictions could be experimentally tested by observing whether the ``threshold'' number of initial mutators changes with $N$.  Here, we have defined the threshold as the number of mutators for which $P_{fix} = 1/2$, and depicted these values with vertical dotted lines.
Parameters are $\alpha=.4,\delta=0,\mu_+=s=1/120$.} 
\label{pfix_vs_n}
\end{figure}

\textbf{Relation to Previous Theoretical Work:}  As mentioned in the introduction, there are many existing theoretical models of mutator evolution.  In this section we briefly review the existing body of knowledge and place our present work in this larger context.  Studies are discussed roughly in order of increasing similarity to our present work. 

\textit{Models with explicit environmental change:} \cite{leighorig} endeavored to calculate the mutation rate that maximizes the growth rate of its corresponding modifier locus.  An infinite population with this wild-type (``resident'') mutation rate is evolutionarily stable in the sense that it cannot be invaded and swept by any modifier of mutation rate.  Such an evolutionarily stable strategy (ESS) is referred to as the ESS mutation rate. \cite{leighorig} developed a simple two locus, two allele model of mutator dynamics in an environment that regularly alternates between two states.  One locus is under selection, and its two alleles are alternately favored in the two different environments.  The second locus is not under direct selection and merely modifies the mutation rate at the selective locus.  The dynamics of the mutator allele are deterministically governed by two effects.  First, immediately after the environment changes, the mutator increases its frequency because the small population of mutants, which is favored in the new environment, is over-represented in the mutator background.  This favors the higher mutation rate.  Secondly, after the mutant sweeps through the population, the frequency of the mutator decreases due to association with the deleterious mutants that it generates at its new fitness peak.  This favors lower mutation rate.  The cycle repeats itself many times, and \cite{leighorig} finds that the long term ESS mutation rate is equal to the rate of environmental change.  Over the years, this basic model was improved by incorporating the effects of timing of environmental changes, varying selective coefficients \citep{ishii}, intermediate genotypes \citep{travis}, and multiple mutable sites \citep{palmer2006iha}.  

While these models doubtless provide valuable insight into certain biological scenarios, they are rather orthogonal to our work.  Three differences seem especially important.  First, most obviously, mutator success requires repeated environmental changes in these models.  In contrast, our model shows that environmental change is only \textit{necessary} for mutator fixation insofar that it provides a rationale for having a population displaced from its fitness peak.  Secondly, they endeavor to find the global ESS mutation rate whereas we focus on quantifying, via fixation probability, the probabilistic result of a single competition experiment.  While full knowledge of $P_{fix}(N,s,\alpha,\mu_+,\mu_-,\delta)$ implies the value of the ESS, the converse is not true.  Thirdly, their mechanism of mutator success is very different from ours.  Whereas they rely upon the alternating selective effects of existing mutants to boost mutator frequency, our model analyzes the dynamic, stochastic interplay between random drift, deleterious mutations, and advantageous mutations in a constant environment.  We propose that, on the whole, our model contains fewer special assumptions than models with explicit environmental change.  Regardless of whether fluctuating or constant environments are more biologically informative, our results constitute an important null model of mutator fixation.

\textit{Constant environment models:}
Work by \cite{tanaka2003emg} also involves a changing environment.  However, unlike the models described in the previous section, theirs contains no alternating selective effects: when the environment changes, the mutations acquired during the previous environmental cycle simply become neutral.  Thus, as in our work, all beneficial mutants are generated \textit{de novo}.  In further similarity with our work, \cite{tanaka2003emg} pursue, via quasi-stochastic simulations and analytic approximations, an understanding of the long term mutator behavior by concentrating on a single environmental cycle, i.e. by examining populations in a constant environment.   These authors were interested primarily in the case when $Nx_o\mu_+ \ll N(1-x_o)\mu_-$, where the fixation of mutators is in some sense unlikely.  With this in mind, instead of $P_{fix}$, they measure and calculate the (much larger) probability $P_{gain}$ that the initially rare mutator increases its frequency by the end of a ``time cycle.''  These cycles are defined to end when an expanding clone in a wild-type background reaches a size of $O(N)$, at which point the simulation is halted. Their most interesting result is that $P_{gain}$ is substantial even when $Nx_o\mu_+ \ll N(1-x_o)\mu_-$.  In other words, mutators can still ``break even'' if the wild-type background generates the first beneficial mutation, which is important if the environment changes.  Nonetheless, without environmental change in their model, mutators will always be doomed unless they are the first to generate a beneficial mutation.  Furthermore, they model birth and death processes deterministically, in a way that precludes extinction.  For these reasons, our $P_{fix}$ and their $P_{gain}$ are truly distinct quantities, and no direct comparison can be made with our work.  

We next discuss a simple calculation by \cite{lenski2004pag} based on indirect mutation-selection equilibrium of the mutator subpopulation.  If the dominant processes occurring in the population are mutation into the mutator state and creation of deleterious mutations by mutators, then the frequency of mutators approaches an equilibrium value.  This frequency is easily calculated if, as in A3 of ISLA, deleterious mutations are treated as immediately lethal:
\begin{eqnarray}
\nonumber x_{eq} = \frac{U}{(1-\alpha_e)(\mu_+ - \mu_-)} \approx \frac{U}{\mu_+(1-\alpha_e)}
\end{eqnarray} 
The time taken for the population to reach this equilibrium state, as well as a much more careful calculation of $x_{eq}$, was investigated by \cite{johnson_prs}, but presently we assume that this simple estimate is sufficient.  In equilibrium, beneficial mutations therefore arise at a rate $Nx_{eq}\mu_+ \alpha_e$ from the mutators, and rate $N(1-x_{eq})\mu_- \alpha_e$ from the wild-type.  If all beneficial mutants of equal effect have the same probability of achieving fixation, regardless of whether they originate in a mutator or wild-type background, then the \textit{fraction of substitutions} linked to a mutator is approximately
\begin{eqnarray}
\frac{U}{\mu_+(1-\alpha_e)} \frac{\mu_+}{\bar \mu} = \frac{U}{\bar \mu (1-\alpha_e)}
\label{lenski_frac}
\end{eqnarray}
Plugging in reasonable values, \cite{lenski2004pag} finds that $\approx 1\%$ of substitutions should be linked to mutators.  Furthermore, given that each line of \textit{E. coli} in experiments by \cite{sniegowski1997ehm} generated 10-20 substitutions, this calculation is impressively consistent with the observation that $3/12$ lines became mutators. 

In order to relate this approach to our own, we must reintroduce dynamics into the picture.  We can interpret the quantity $x \frac{\mu_+}{\bar \mu}$ as the conditional probability that a mutator achieves fixation, given that a selective sweep occurs during its lifetime.  Our quantity $P_{fix}$ is this conditional probability multiplied by the probability that a selective sweep occurs during the lifetime of a mutator.  Assuming that selective sweeps and death each occur as Poisson processes with rates $N\bar \mu \alpha_e s$ and $(\mu_+-\mu_-)(1-\alpha_e)$, respectively, it is straightforward to show that the probability that at least one selective sweep occurs before death is given by
\begin{eqnarray}
\nonumber \frac{N\bar \mu \alpha_e s}{(1-\alpha_e)(\mu_+ - \mu_-)} \left(1+ \frac{N\bar \mu \alpha_e s}{(1-\alpha_e)(\mu_+ - \mu_-)}\right)^{-1} 
\end{eqnarray}
Multiplying this expression by the conditional probability $x \frac{\mu_+}{\bar \mu} \approx x R$, we obtain Eq.\ref{asymp_mum}.  Thus, the approach suggested by \cite{lenski2004pag} is the equilibrium version of ISLA, in the limit where mutational processes occur frequently enough to overwhelm random genetic drift.  Thus, remarkably, even though this approach frames the problem of mutator fixation in terms of competition with beneficial mutations in wild-type backgrounds, $R$ cancels out of the solution in the strong-effect mutator regime: $R \gg N \alpha_e s/(1-\alpha_e)$.

It is also worthwhile to examine the conditions under which we expect the equilibrium assumption to hold.  Let us imagine that an evolution experiment is conducted for $T$ generations, during which $H$ substitutions occur.  ISLA predicts that the expected number of mutator fixations is $N P_{fix}  U T$, whereas according to Eq.\ref{lenski_frac}, the equilibrium approach yields a value equal to $H \frac{U}{\bar \mu}$.
Setting these two values equal to one another, and plugging in (from Eq.\ref{s_max}) $P_{fix}(x_o=1/N) = s \frac{\alpha_e}{1-\alpha_e}$,  we obtain 
\begin{eqnarray}
\nonumber H = N s \bar \mu T \frac{\alpha_e}{1-\alpha_e} \approx N \bar \mu \alpha_e s  T 
\end{eqnarray}
This expression merely states that the (mostly wild-type) population is in the ``successive mutations regime'', i.e. only a single beneficial mutation spreads at a time.  Alternatively, one could imagine turning this argument around and asking what $P_{fix}$ must equal given that the equilibrium approach is valid and that the population accumulates substitutions ``one by one''.  In that case, one would, remarkably, arrive at $P_{fix}(x_o=1/N) = \alpha_e s$, which (for small $\alpha_e$ and $NS_\mu \gg 1$) is what we obtained earlier (Eq.\ref{s_max}) by more sophisticated methods.  

Turning to another study, \cite{tenaillon1999mps} investigated, via stochastic simulations and very brief analytic arguments, multi-locus mutator evolution in a constant environment.  These extensive simulations are a generalization of earlier work by \cite{taddei1997rma} and are partly amenable to comparison with our work.  Some noteworthy differences with our simulations are that they scan a larger range of $N$, they have a more realistic implementation of mutation, and, most importantly, they allow flux into and out of the mutator state.  Thus, mutators are never absolutely fixed during their trials, which necessitates a different termination condition than ours:  They declare a trial ``over'' when the population reaches its maximum fitness, whereas we declare it ``over'' when the mutator is completely and permanently fixed or lost.  Upon termination of the trial, they consider the mutator ``fixed'' if its frequency is $> 95\%$.  They measure the fraction of trials that terminate with mutator frequency $> 95\%$ and denote this quantity the ``frequency of mutator fixation,'' which differs from our $P_{fix}$ because of reasons discussed below.  

One important consequence of their method is that the total number of mutators \textit{generated} during a trial varies with the choice of parameters.  This is because each replication event presents a chance for the creation of a new mutator, and the number of replication events that occur before termination clearly depends on $N$, $s$, $\mu_+, \mu_-$, and the number of mutational steps required to reach the peak.  Thus, a change in the value of any of these parameters may alter the  ``frequency of mutator fixation'' simply because it changes the number of mutators that are typically created during the trial.  Our $P_{fix}$, on the other hand, remains invariant under such changes and allows us to filter out this background effect.  Their system is doubtless a more literally accurate representation of biological reality, which has its virtues but also major costs, which we discuss below in the context of two important examples.

First, they measure that the ``frequency of mutator fixation'' increases with $N$.  This is an interesting and potentially practical result, but their method makes it very difficult to determine the extent to which the increase is simply due the background effect that more mutators were created in the larger populations.  ISLA, on the other hand, unambiguously states that when $NS_\mu \gg 1, P_{fix}$ for a single mutator becomes independent of $N$.  Therefore, ISLA predicts that the dependence of mutator fixation frequency on population size observed by \cite{tenaillon1999mps} is entirely driven by the simple background effect.  

A second example has even more dramatic conceptual consequences.   These authors ask whether $P_{fix}$ is determined by the number of potentially advantageous mutations (steps away from the peak) or merely by the \textit{rate} that such mutations are generated.  In order to investigate this question, they devised two sets of simulations.  In one set, there were 12 available advantageous mutations, accessible at a rate of $10^{-8}$ each.  In the other set, there was a single mutation of the same effect, accessible at a rate of $12 \times 10^{-8}$.  The explicit difference between these sets of simulations is the number of steps to the fitness peak, but an additional, implicit difference is that the set with 12 beneficial mutations runs for more generations.  Therefore, more mutators are created in that set of simulations.  Now, ISLA predicts that $P_{fix}$ depends only on the  advantageous mutation \textit{rate}, and that therefore the two simulations should result in the same $P_{fix}$.  In seeming contrast, they found the ``frequency of mutator fixation'' to equal approximately $.5$ for the first situation and approximately zero for the second.  This observation led them to conclude that mutators succeed because of their advantage in rapidly creating genomes which carry multiple beneficial mutations, which is fundamentally different from our conceptual picture.  We propose that this simulation finding might be explained by the simple background effect that far more mutators are created en route to acquiring $12$ beneficial mutations than to acquiring a single beneficial mutation.  ISLA completely neglects multiple beneficial mutations, and its success, both near the peak (Fig.\ref{trans_fig}) and far from it (Fig.\ref{mod_sim_comp_fig}), suggests that the multiple mutations effect proposed by \cite{tenaillon1999mps} in  fact plays a very minor role in mutator fixation.  However, it should be noted that we did not investigate cases where the mutator is \textit{favored} and only a single beneficial mutant is available.  It could be the case that multiple beneficial mutations in the same genome are implicitly important in that they are what allows the mutator to overcome competition with wild-type beneficial mutations. This hypothesis should be explored in future work.  

Whereas \cite{tenaillon1999mps} focused almost exclusively on stochastic simulations, work by \cite{andre2006emr} relies almost exclusively on analytic methods.  In work that bears many similarities to ours, \cite{andre2006emr} studied, mostly via an analytic approach, the long term trajectory of mutation rate evolution.  A key insight of theirs is that, in a finite asexual population, the frequency of a mutator undergoes strong fluctuations, with values covering the entire range from zero initially to one upon a selective sweep by a linked locus.  Thus, they point out that studies which assume that mutators are rare during all generations, either because of infinite population size \citep{leighorig} or sexual recombination \citep{johnson1999bmh}, are qualitatively different than finite asexual populations.  \cite{andre2006emr} remedy this problem by calculating the \textit{fixation probability} of an initially rare mutator.  We now briefly summarize their method of solution and show that, with minor modification, it corresponds to the $N \mu \rightarrow \infty$ limit of our results.  In what follows, we take some liberty in changing their notation and using continuous time.  

Their initial condition is identical to ours: a clonal population is seeded with a small number of otherwise identical mutators.  They then temporarily ignore beneficial mutations and analyze how the \textit{expected} number of mutators changes with time.  In agreement with \cite{johnson_prs}, they find that after a waiting time $1/s$, the mutator subpopulation declines exponentially, i.e. $E[x(t)] = x_oe^{-(\mu_+-\mu_-)(1-\alpha_e)(t-1/s)}$.  They then construct their key equations (their Eq. 19)
\begin{eqnarray}
\nonumber \frac{d}{dt}P_{fix}(t) &=& (1-P_{fix}(t)-P_{loss}(t))\cdot N\bar{\mu}\alpha_e s \cdot \frac{\mu_+}{\bar{\mu}} \cdot E[x(t)] \\
\nonumber \frac{d}{dt}P_{loss}(t) &=& (1-P_{fix}(t)-P_{loss}(t))\cdot N\bar{\mu}\alpha_e s \cdot \frac{\mu_-}{\bar{\mu}} \cdot (1-E[x(t)])
\end{eqnarray}
We have written these equations in a somewhat peculiar way, and replaced their symbol $K$ with $N\bar{\mu}\alpha_e s$ in order to facilitate translating between our notation and theirs.  These equations are very similar to ISLA in that they represent the instantaneous fixation of beneficial mutations which originate from a time dependent mutator subpopulation.  However, there are two disturbing features about these equations.  First, they assume that the only cause of mutator extinction is beneficial mutations in the wild-type background.   In fact, mutators also become extinct due to (i) their mutational load and (ii) random drift.  In their equations, $E[x(t)]$ declines exponentially, but erroneously, this decline does not contribute to $P_{loss}$.  Both (i) and (ii) cause an overestimate of $P_{fix}$.  The second disturbing feature of these equations is the appearance of expectation values on the RHS.  With this move, \cite{andre2006emr} replaced the random variable $x(t)$ with its mean value, which is a very substantive approximation.  The distribution of $x(t)$ is in fact diffusing, i.e. random drift is in fact occurring.  Nevertheless, we expect that their representation of $x(t)$ as a deterministic quantity to be approximately valid when the timescale of this diffusion is slower than the timescales due to mutation and selection.  Unlike our approach, theirs cannot quantify when it is safe to neglect random drift.  Looking back to Eq.\ref{back_eq}, we see that the diffusive process, i.e. random drift, can be neglected when $N \mu_\pm (1-\alpha_e)  \gg 1$ and $N^2\mu_\pm \alpha_e s \gg 1$.  It just so happens that these criteria will often be met in microbial populations.  

We now explicitly demonstrate some important parallels between our work and that of \cite{andre2006emr} in the large $N \mu$ limit.  Since, in our model, deleterious mutations are as strong as advantageous ones, the best comparison is made with their ``ruby in the rubbish'' hypothesis.  The relevant solution is their Eq.A5
\begin{eqnarray}
P_{fix} = x_o \frac{N \bar{\mu} \alpha_e s}{1-(1-N\bar{\mu} \alpha_e s)\cdot e^{-(\mu_+-\mu_-)(1-\alpha_e)}}
\cdot \frac{\mu_+}{\bar{\mu}}
\label{ag_pfix}
\end{eqnarray}
Simplifying the denominator by taking $\exp\left[-(\mu_+ - \mu_-)(1-\alpha_e)\right] \approx 1-(\mu_+-\mu_-)(1-\alpha_e)$ and neglecting the term $-N\bar \mu \alpha_e s \mu_+$, we recover our large $N \mu$ result from ISLA (Eq.\ref{asymp_mum}).  The neglected term inflates the value of $P_{fix}$, and is a result of these authors not treating extinction of the mutator due to its mutational load.  This has important consequences for the next topic. 

\textbf{Long term mutation rate evolution:}  Although our work primarily addresses the plain issue of calculating $P_{fix}$, we briefly contemplate implications for the more grand question of long term mutation rate evolution.

\textit{$\mu_{conv}$ is proportional to the rate of sweeps:}
Thus far we have considered selective sweeps to be initiated by \textit{de novo} beneficial mutations.  Let us now briefly apply our results to the case where sweeps are instead triggered by an environment that changes at rate $K$.  This merely requires transcribing $N \bar{\mu} \alpha_e s \leftrightarrow K$.  Following \cite{andre2006emr} we expand the fixation probability (Eq.\ref{ag_pfix}) in powers of $\mu_+ - \mu_-$ and denote the first order coefficient in this series by $Sel(\mu_-)$.  The roots of $Sel(\mu_-)$ give the ``convergence stable resident mutation rate.''  Using Eq.\ref{asymp_mum}, we find $\mu_{conv} = K/(1-\alpha_e) \approx K$, which is the classical result \citep{leighorig}.  Using Eq.\ref{ag_pfix}, \cite{andre2006emr} find a qualitatively different result: $\mu_{conv} =\frac{K}{(1-\alpha_e)(1-K)}$, which diverges as $K \rightarrow 1$.  The reason for this discrepancy is that \cite{andre2006emr} did not allow for extinction due to the mutational load.  ISLA naturally accounts for these extinction events and leads to the classical result.  However, ISLA approximates deleterious mutations as being lethal, whereas these authors also treated the more realistic non-lethal case.  It may be possible to demonstrate, via further analysis, the claim that non-lethal deleterious mutants cause $\mu_{conv}$ to diverge for some parameter values.

\textit{Equilibrium mutation rate:}
We find that $Sel(\mu_-) = \frac{1}{\mu_-}\frac{\alpha_e(Ns+1)-1}{N\alpha_e s}$, whereas \cite{andre2006emr} find $Sel(\mu_-) =  \frac{1}{\mu_-}\frac{\alpha_e(Ns+1)-1}{N\alpha_e s} + 1-\alpha_e$.  Our expression indicates that there are no equilibrium mutation rates: for all $\mu_-$, weak mutators are favored when $\alpha_e(Ns+1) \approx N\alpha_e s > 1$ and disfavored in the opposite case.  This threshold is clearly in agreement with our Eq.\ref{transition}.  Thus, as far as ISLA is concerned, populations with $N \alpha_e s < 1$ should continually evolve toward the minimum attainable mutation rate.  On the other hand, populations with $N \alpha_e s > 1$ should evolve an ever higher mutation rate.  Our expression for $Sel(\mu_-)$ is clearly inaccurate for very small $\mu_-$ (because random drift dominates in that regime) and also for very large $\mu_-$ (since our simulations show that there is a maximum mutation rate that can achieve fixation).  

\textbf{Limitations of Present Work:} Real biological populations possess many features that this article either neglects or severely constrains.  We now briefly discuss the most striking limitations.  

\textit{Initial Conditions:}  Both ISLA and our simulations suppose that ``initially'' all members of the population have the same fitness.  If this assumption is false and mutators arise randomly in a population with pre-existing fitness variation, this might act to decrease mutator success: unless the mutator happens to emerge from the fittest subclass of the population, the advantageous mutations it generates will already be present in more abundant subclasses which could out-compete the rare mutator.  This point is especially relevant since, in comparing ISLA to experiment, we essentially assumed that each mutator that arose during the course of the experiments did so in a population consisting of a single fitness value.   

\textit{Strict Asexuality:} Our simulations and ISLA do not allow any mechanisms of horizontal gene transfer or recombination.  These events would decouple mutator alleles from the advantageous mutations that they generated, and thereby result in significantly decreased mutator success.  This effect is especially important since some genes associated with a mutator phenotype also exhibit hyper-recombination \citep{maticrev}.

\textit{Simple Fitness Landscape:}  Our simulations assume that mutations all fall into one of three classes: lethal, beneficial with effect $+s$, or deleterious with effect $-s$.   As mentioned previously, and discussed in Supplementary Information, it may be true that, in large populations, beneficial mutations of a fixed size $\tilde{s}$ are the ones that typically reach appreciable frequency \citep{gerrish1998fcb,d_fisher_short,hegreness2006epi}.  However, this simplification is certainly not possible when considering deleterious mutants, whose distribution is likely complicated and bimodal, with many mutations being nearly neutral and many being lethal \citep{eyrewalker2007dfe}.  Fig.\ref{lethal_comp} suggests that increasing the strength of deleterious mutations has effects only at large $\mu_+/s$, where it increases both the peak value of $S_\mu$ and the value $\mu_+/s$ at which the peak occurs.  Along these lines, a simulation model that included a class of weakly deleterious mutations would likely continue this trend.  This would delay the large discrepancy between the simulations and ISLA until even larger $\mu_+/s$.  This issue could help to explain the previously mentioned fact that $\mu_+$ in experiments of \cite{sniegowski1997ehm} seem very close to the maximum allowable value.  Including mildly deleterious mutations would also prolong the lifetime of genomes which carry them.  In this case, it might be necessary to incorporate a time delay before these deleterious mutations are ``enforced,'' along the lines explored by \cite{johnson_prs}.

\textbf{Suggestions for further research:}  This article leaves many questions unanswered, but also points to interesting theoretical and experimental opportunities.  

\textit{Theoretical directions:} A satisfactory analytic description of our stochastic simulations remains incomplete.  Two key issues remain unresolved.  First, we do not understand the mechanism by which mutators continue to succeed when faced with intense mutational competition from the wild-type background (Fig.\ref{mum_comp_fig}).  Our work and that of \cite{andre2006emr} both imply that mutations in wild-type backgrounds should become important when $N \alpha_e s \sim \mu_+/\mu_-$, but this is not borne out in the simulations unless $s$ is ``sufficiently large.''  Secondly, it is clear that ISLA fails to match simulations when the mutation rate is very large $(1-\alpha_e)\mu_+ \gtrsim s$.  Quantifying the success of mutators in this regime is especially relevant to studies of long term mutation rate evolution.  

Another issue that we did not address is the full dynamics of mutator fixation.  Our analytic results are mostly derived from Eq.\ref{back_eq}, which is relevant to the eventual fate of mutators.  An approximate solution to the time dependent forward diffusion (Eq.\ref{full_fpe}), with $\mu_- = 0$, is given in Supplementary Information.  This solution provides some dynamical information, but, like the entire ISLA approach, it assumes that selective sweeps occur instantaneously.  In this sense, Eq.\ref{full_fpe} predicts incorrect dynamics.  Furthermore, we showed that mutator success is compactly represented by an effective selection coefficient $S_\mu$.  For simple advantageous mutants, $S$ contains information not only about $P_{fix}$ but also about the average dynamics: $\langle x(t) \rangle \sim e^{St}$ when rare.  Perhaps that is the case with mutators as well. 

\textit{Experimental ideas:} Our work shows that, in most regimes, $P_{fix}$ is not explicitly frequency dependent.  Rather, $P_{fix}$ depends on the initial \textit{number} of mutants $Nx_o$.  This scaling behavior could be tested experimentally.  Suppose that competition experiments in a chemostat carrying a population of size $N_1$ showed that, when the initial frequency of mutators exceeded a threshold value of $x_1$, mutator achieved fixation with a high probability.  One could decrease the population size to $N_2$ and again inoculate with mutators at a frequency of $x_1$.  Our results predict that mutators would not achieve fixation in this case because $N_2 x_1$ is less than the threshold number $N_1 x_1$.  In fact, very similar experiments were recently performed by \cite{lechat2006ecm}, which support the notion that $P_{fix}$ scales with $Nx_o$ and not with $x_o$ alone.  However, these competition experiments were done under a lethal selective pressure, which selected for pre-exiting resistant mutants.  Here we propose competitions between initially isogenic (aside from the mutator allele) mutator and wild-type strains adapting to a new environment.  In addition to this scaling behavior, ISLA predicts a testable value for this threshold that differs significantly from the frequency dependent picture represented by Eq.\ref{freq_dependent}.  These ideas are presented in Fig.\ref{pfix_vs_n}.

It would also be interesting to experimentally investigate the decline in mutator success seen for very large mutation rates when $(1-\alpha_e)\mu_+ \sim s$.  As mentioned previously, during the first few thousand generations of experiments by \cite{sniegowski1997ehm}, $N\alpha_e s \approx 1.1$.  The reason why no mutators achieved fixation after the first $10,000$ generations could be that this parameter decreased below the threshold value of one during the course of its evolution.  A similar effect was previously discussed by \cite{kessler1998mds}.  An alternative explanation is that $\mu_+$ was near the theoretical maximum $(1-\alpha_e)\mu_+ \sim 1$ suggested from our simulations.  As noted by \cite{gerrish2007cgl}, once could test these competing explanations by founding several new lineages with a clone from of one of the mutator populations, and growing these mutator lineages in a novel environment.  The new environment should be one in which $N\alpha_e s > 1$.  If no ``double mutators'' arose, then the hypothesis of a maximum allowable mutation rate would be supported.  
\\

\normalsize We thank Terry Hwa, Ulrich Gerland, Lin Chao, and Aaron Trout for helpful discussions. We also thank two anonymous reviewers for their thoughtful and constructive criticism. The work of HL and CSW was supported in part by the NSF PFC-sponsored Center for Theoretical Biological Physics (Grant No. PHY-0822283).
The work of DAK was supported in part by the Israel Science Foundation.
\bibliographystyle{genetics}
\bibliography{mut_bib.bib}

\end{document}